\documentclass[
12pt,
preprint,preprintnumbers,nofootinbib,
groupedaddress,superscriptaddress,amsmath,amssymb]{revtex4}
\usepackage{graphicx}
\usepackage{dcolumn}
\usepackage{bm}
\usepackage{amssymb}
\usepackage{amsmath}
\usepackage{epsfig}    
\usepackage{color}
\usepackage{hhline}

\def\be{\begin{equation}}
\def\ee{\end{equation}}
\newcommand{\bea}{\begin{eqnarray}}
\newcommand{\eea}{\end{eqnarray}}
\newcommand{\nn}{\nonumber}

\numberwithin{equation}{section}

\begin{document}

\title{A radiative seesaw in a non-holomorphic modular $S_3$ flavor symmetry}

\author{Hiroshi Okada}
\email{hiroshi3okada@htu.edu.cn}
\affiliation{Department of Physics, Henan Normal University, Xinxiang 453007, China}

\author{Yuta Orikasa}
\email{Yuta.Orikasa@utef.cvut.cz}
\affiliation{Institute of Experimental and Applied Physics, 
Czech Technical University in Prague, 
Husova 240/5, 110 00 Prague 1, Czech Republic}

\date{\today}

\begin{abstract}
We study a non-holomorphic modular $S_3$ flavor symmetry in which we analyze neutrino sector, dark matter, and lepton flavor violations.
The active neutrino mass is generated via one-loop level.
We achieve chi-square analysis and demonstrate some predictions in cases of normal hierarchy with fermionic or bosonic dark matter and inverted hierarchy of fermionic or bosonic dark matter.

\end{abstract}
\maketitle
\newpage

\section{Introduction}
Neutrino physics still possesses unknown issues such as tiny neutrino masses, phases and mixings very well, which would be explained by beyond the standard model (BSM).
To understand the tiny neutrino masses, a radiative seesaw model is one of the promising candidates since the neutrino mass
is induced at loop-levels. This kind of model is renowned as a TeV scale scenario which would reach current experiments such as large colliders.
In addition, we could include a dark matter (DM) candidate that interacts with neutrinos.
In order to resolve the phases and mixings, on the other hand, flavor symmetries plays an important role and would predict some of the observables that would be recovered via future experiments.
In particular, non-Abelian discrete symmetries are one of the good candidates to predict these observables and a vast literature has been arisen in the last several decades.~\footnote{These symmetries have initially been applied via multi-Higgs scenarios(flavons) and useful reviews are found in refs.~\cite{Altarelli:2010gt, Ishimori:2010au, Kobayashi:2022moq, Hernandez:2012ra, King:2013eh, King:2014nza, King:2017guk, Petcov:2017ggy, Ishimori:2012zz}.}
 
In our paper, we apply a one-loop induced neutrino mass model based on a scenario in ref.~\cite{Ma:2006km} for the lepton sector in order to explain the tiny neutrino masses and DM candidate. Then, we discuss lepton flavor violations (LFVs) which would be regions nearby the upper bounds of current experiments. 
 In addition, we adopt a non-holomorphic modular $S_3$ flavor symmetry to explain the phases and mixings.
 \footnote{A modular $S_3$ symmetry is applied for lepton and quark sector in refs.~\cite{Kobayashi:2018vbk, Kobayashi:2018wkl, Kobayashi:2019rzp, Okada:2019xqk, Mishra:2020gxg, Du:2020ylx, Mishra:2024fcr, Behera:2024ark, Behera:2024vfv, Meloni:2023aru}, but  the most of papers have been worked on holomorphic scenarios such as supersymmetric theory that is initially proposed by ''Feruglio''~\cite{Feruglio:2017spp}.
The non-holomorphic modular symmetries are recently developed by "Qu" and "Ding"~\cite{Qu:2024rns}
which is one of the greatest breakthroughs to apply for all the non-holomorphic scenarios of BSM.~
After the paper, several ideas are released in refs.~\cite{Nomura:2024atp, Ding:2024inn, Li:2024svh, Nomura:2024nwh, Nomura:2024vzw}.}
Then, we analyze our model with chi-square analysis to fit the neutrino oscillation data of Nufit6.0~\cite{Esteban:2024eli} and satisfy the LFVs.
Then we discuss the DM candidate in both cases of a fermion and a boson depending on the normal hierarchy (NH) or inverted hierarchy (IH). 
Finally, we demonstrate each case of prediction for the neutrino sector and DM mass.

This paper is organized as follows.
In Sec.~\ref{sec:realization},   we give our model set up under the non-holomorphic modular $S_3$ flavor symmetry writing down Higgs potential and each of mass matrices. 
Then, we formulate the LFVs and neutrino sector.
In Sec. III, , we discuss our DM candidate; fermionic (FDM) and bosonic one (BDM), in which we show the main interactions to explain the correct relic density considering direct detection searches.
In Sec.~IV, we show predictions for each case of FDM and BDM of NH, and   FDM and BDM of IH.
Finally we conclude and discuss in Sec.~\ref{sec:conclusion}.

\section{Model} 
\label{sec:realization}

In this section, we review our model.
Our field contents are the same as the Ma model~\cite{Ma:2006km}.
We assign singlet and doublet representations under $S_3$ to left-handed leptons $\overline{L_{L_e}}, \ \overline{L_{L_2}}\equiv(\overline{ L_{L_\mu}},\overline{ L_{L_\tau}})^T$ and right-handed charged-leptons $e_{R_e},\ 
e_{R_2}\equiv(e_{R_{\mu}},e_{R_{\tau}})^T$, respectively. In addition, modular weights are respectively assigned by $-2$ and $0$ for left-handed and right-handed ones. The right-handed neutrinos are assigned by a doublet under $S_3$ where the zero modular weight is assigned.
Due to two right-handed neutral fermions, we have two mass eigenstates for our active neutrinos, as will be discussed later.
The SM Higgs is denoted by $H\equiv (h^+, v_H+h+iz)^T/\sqrt2$ where $v_H \approx 246$ GeV, $h^+$ and $z$ are respectively replaced by the massive SM gauge bosons $W^+$ and $Z$. $h$ is the SM Higgs  and its mass is $m_h\approx126$ GeV.
Inert Higgs $\eta \equiv (\eta_R+i\eta_I, \eta^-)^T/\sqrt2$ has $+1$ modular weight and its hypercharge is $-1/2$ instead of $+1/2$.
All valid fields and their assignments are summarized in Tab.~\ref{tab:fields}.
 

\begin{center} 
\begin{table}[tb]
\begin{tabular}{|c||c|c|c|c|c||c|c||}\hline\hline  
&\multicolumn{5}{c||}{ Fermions} & \multicolumn{2}{c||}{Bosons} \\\hline
  & ~$\overline{L_{L_e}}$~& ~$\overline{L_{L_2}}\equiv(\overline{ L_{L_\mu}},\overline{ L_{L_\tau}})^T$~ & ~$e_{R_e}$~& ~$e_{R_2}\equiv(e_{R_{\mu}},e_{R_{\tau}})^T$~ & ~$N_{R}$~ & ~$H$~  & ~$\eta$~
  \\\hline 
 $SU(2)_L$ & $\bm{2}$  & $\bm{2}$  & $\bm{1}$    & $\bm{1}$  & $\bm{1}$ & $\bm{2}$ & $\bm{2}$    \\\hline 
$U(1)_Y$ & $\frac12$ & $\frac12$ & $-1$  & $-1$& $0$   & $\frac12$  & -$\frac12$      \\\hline
 $S_3$ & $1$ & $2$ & $1$ & $2$ & $2$  & $1$ & $1$     \\\hline
 $-k$    & $-2$ & $-2$ &$0$ & $0$ & $+1$   & $0$ & $+1$   \\\hline
\end{tabular}
\caption{Field contents of fermions and bosons
and their charge assignments under $SU(2)_L\times U(1)_Y\times S_{3}$ in the lepton and boson sector, 
where $-k$ is the number of modular weight 
and the quark sector is the same as the SM.}
\label{tab:fields}
\end{table}
\end{center}

{\it The invariant Higgs potential} under these symmetries is given by
 is given by
\begin{align}
{\cal V} &= -\mu_H^2 |H|^2 +\mu^2_\eta |\eta|^2\\
&+ \frac14 \lambda_H|H|^4+ \frac14\lambda_\eta |\eta|^4
+\lambda_{H\eta} |H|^2|\eta|^2+\lambda_{H\eta}'  |H \eta|^2
+\frac14[ \lambda_{H\eta}''  (H \eta)^2+ {\rm h.c.}]\nn,
 \label{eq:pot}
\end{align}
that is exactly the same as the potential in Ma model and  any couplings $\mu_\eta^2$, $\lambda_{H\eta},\ \lambda'_{H\eta}$ include $(i\tau-i\bar \tau)$, $\lambda_\eta$ includes $(i\tau-i\bar \tau)^2$, and $\lambda''_{H\eta}$ includes $Y^{(-2)}_{\bf 1}$, in order to be invariant under the modular $S_3$ symmetry. Thus, we obtain the same mass eigenvalues as the Ma model.

{\it The charged-lepton mass  matrix} under these symmetries is given after spontaneous symmetry breaking of the following Lagrangian:
\begin{align}
&\beta_\ell[(\overline{L_{L_\mu}} e_{R_\mu}-\overline{L_{L_\tau}} e_{R_\tau})y_1-
(\overline{L_{L_\mu}} e_{R_\tau} +\overline{L_{L_\tau}} e_{R_\mu})y_2] H \nn\\
&+\delta_\ell(y_1\overline{L_{L_\mu}}+y_2\overline{L_{L_\tau}})e_{R_e} H
+\gamma_\ell \overline{L_{L_e}}(y_1 e_{R_\mu}+y_2e_{R_\tau}) H ,
\end{align}
and its form is found as
\begin{align}
m_\ell&= \frac {v_H}{\sqrt{2}}
\left[\begin{array}{ccc}
0 & \gamma_\ell y_1 & \gamma_\ell y_2 \\ 
\delta_\ell y_1 & \beta_\ell y_1 &  -\beta_\ell y_2 \\ 
\delta_\ell y_2 & -\beta_\ell y_2 &  -\beta_\ell y_1 \\ 
\end{array}\right], 
\end{align}
where {$\langle H\rangle\equiv (0, v_H/\sqrt2)^T$} and $Y^{(2)}_{\bf 2}\equiv (y_1,y_2)^T$.
Then the charged-lepton mass eigenstate can be found by $|D_\ell|^2\equiv V_{L} m_\ell m^\dag_\ell V_{L}^\dag$.
We numerically fix the free parameters $\beta_\ell,\gamma_\ell,\delta_\ell$, which are real without loss of generality, to fit the three charged-lepton mass eigenvalues.

{\it  The right-handed neutrino mass matrix} under these symmetries
arises from the following Lagrangian
\begin{align}
&M_1(\overline{N^c_{R_1}}N_{R_1} +\overline{N^c_{R_2}}N_{R_2}) \nn\\
&+M_2[(\overline{N^c_{R_1}}N_{R_1} -\overline{N^c_{R_2}}N_{R_2})y'_1
-
(\overline{N^c_{R_1}}N_{R_2} +\overline{N^c_{R_2}}N_{R_1})y'_2],
\end{align}
where $M_1$ includes a trivial singlet modular form with $-2$ weight. 
Its form is given by
\begin{align}
{\cal M_N} &=M_2
\left[\begin{array}{cc}
\tilde M_1 +y'_1 & -y'_2 \\ 
-y'_2 & \tilde M_1 - y'_1   \\ 
\end{array}\right]
\equiv M_2 \tilde{\cal M}_{\cal N},
\end{align}
where $Y^{(-2)}_{\bf 2}\equiv (y'_1,y'_2)^T$.
${\cal M_N} $ is diagonalized by $\tilde D_N=V_N^T \tilde{\cal M}_{\cal N} V_N$.
$M_2$ is real while $\tilde M_1\equiv M_1/M_2$ is a complex mass parameter.

{\it The Dirac Yukawa matrix} under these symmetries is arisen from the following Lagrangian:
\begin{align}
&\gamma_\nu[(\overline{L_{L_\mu}} N_{R_1}-\overline{L_{L_\tau}} e_{R_2})f_1-
(\overline{L_{L_\mu}} N_{R_2} +\overline{L_{L_\tau}} N_{R_1})f_2] \eta \nn\\
&+\alpha_\ell \overline{L_{L_e}}(f_1N_{R_1}+f_2 N_{R_2})  \eta
+\beta_\ell (\overline{L_{L_\mu}} N_{R_1}+ \overline{L_{L_\tau}} N_{R_2}) \eta ,
\end{align}
and its form in basis of $\overline{L_{L}} y_\eta e_R \eta$ is given by
\begin{align}
y_\eta &=\gamma_\nu
\left[\begin{array}{cc}
\tilde \alpha_\nu  f_1 &\tilde \alpha_\nu  f_2 \\ 
\tilde\beta_\nu +  f_1 & - f_2 \\ 
- f_2 &\tilde\beta_\nu -  f_1   \\ 
\end{array}\right]
\equiv \gamma_\nu \tilde y_\eta
,\label{eq:mn}
\end{align}
where $Y^{(0)}_{\bf2}\equiv[f_1,f_2]^T$ and $\tilde \alpha_\nu \equiv \alpha_\nu/\gamma_\nu, \tilde\beta_\nu \equiv \beta_\nu/\gamma_\nu, \gamma_\nu$ are real without loss of generality.

{\it Lepton flavor violations} arises from $y_\eta$ as~\cite{Baek:2016kud, Lindner:2016bgg}
\begin{align}
&{\rm BR}(\ell_i\to\ell_j\gamma)\approx\frac{48 \pi^3\alpha_{em} C_{ij}}{G_F^2 (4\pi)^4}
\frac{\gamma_\nu^4}{ M_2^4 }
\left|\sum_{\alpha=1-3}\tilde Y^\ell_{\eta_{j\alpha}} (\tilde Y^\ell_{\eta_{\alpha i}})^\dag F(\tilde D_{N_{\alpha}} ,\tilde m_{\eta^\pm})\right|^2,\\
&F(m_a,m_b)\approx\frac{2 m^6_a+3m^4_am^2_b-6m^2_am^4_b+m^6_b+12m^4_am^2_b\ln\left(\frac{m_b}{m_a}\right)}{12(m^2_a-m^2_b)^4},
\end{align}
where $Y^\ell_\eta\equiv V^\dag_L \tilde y_\eta V_N$, $C_{21}=1$, $C_{31}=0.1784$, $C_{32}=0.1736$, $\alpha_{em}(m_Z)=1/128.9$, and $G_F=1.166\times10^{-5}$ GeV$^{-2}$.
The experimental upper bounds are given by~\cite{ Davidson:2022nnl}
\begin{align}
{\rm BR}(\mu\to e\gamma)\lesssim 4.2\times10^{-13},\quad 
{\rm BR}(\tau\to e\gamma)\lesssim 3.3\times10^{-8},\quad
{\rm BR}(\tau\to\mu\gamma)\lesssim 4.4\times10^{-8},\label{eq:lfvs-cond}
\end{align}
which will be imposed in our numerical calculation.
 %

{\it Neutrino mass matrix} under these symmetries  is given at one-loop level by
\begin{align}
m_{\nu_{ij}}& = M_2 \gamma_\nu^2 \sum_{\alpha=1,2}
\frac{\tilde Y^\nu_{\eta_{i\alpha}} \tilde D_{N_\alpha} (\tilde Y^\nu_{\eta_{j \alpha}})^T}{(4\pi)^2}
\left(\frac{\tilde m_R^2}{\tilde m_R^2-\tilde D^2_{N_\alpha}} \ln\left[\frac{\tilde m_R^2}{\tilde D^2_{N_\alpha}}\right]
-
\frac{\tilde m_I^2}{\tilde m_I^2-\tilde D^2_{N_\alpha}} \ln\left[\frac{\tilde m_I^2}{\tilde D^2_{N_\alpha}}\right]
\right)\nn\\
&\equiv\kappa \tilde m_{\nu_{ij}} ,
\end{align}
where $\kappa\equiv M_2 \gamma_\nu^2$, $m_{R(I)}\equiv M_2 \tilde m_{R(I)}$ is a mass of the real (imaginary) component of $\eta^0$.
Then the neutrino mass matrix is diagonalized by a unitary matrix $U_{\nu}$ as $U_{\nu}m_\nu U^T_{\nu}=$diag($D_{\nu_1},D_{\nu_2},D_{\nu_3}$). 
 Note here the lightest neutrino mass eigenvalue is zero and we have a Majorana phase matrix which is given by diag.$[1,e^{i\alpha_{21}/2},1]$. 
%
$\sum_{i=1}^3D_{\nu_i}(\equiv \sum D_\nu)$ is constrained by the recent  DESI and CMB data combination~$\sum D_{\nu}\le$ 72 meV~\cite{DESI:2024mwx} as well as the minimal cosmological model $\sum D_{\nu}\le$ 120 meV~\cite{Vagnozzi:2017ovm, Planck:2018vyg}.
 Observed lepton mixing $U_{PMNS}$ is defined by $V^\dag_{eL} U_\nu$.
Each of the mixing angle is given in terms of the component of $U_{PMNS}$ as follows:
\begin{align}
\sin^2\theta_{13}=|(U_{PMNS})_{13}|^2,\quad 
\sin^2\theta_{23}=\frac{|(U_{PMNS})_{23}|^2}{1-|(U_{PMNS})_{13}|^2},\quad 
\sin^2\theta_{12}=\frac{|(U_{PMNS})_{12}|^2}{1-|(U_{PMNS})_{13}|^2}.
\end{align}
The effective mass for the neutrinoless double beta decay is given by
\begin{align}
({\rm NH})&: m_{ee}=|D_{\nu_2} \sin^2\theta_{12} \cos^2\theta_{13}e^{i\alpha_{21}} + D_{\nu_3} \sin^2\theta_{13}e^{-2i\delta_{CP}}|, \\
({\rm IH})&: m_{ee}=|D_{\nu_1} \cos^2\theta_{12} \cos^2\theta_{13}+D_{\nu_2} \sin^2\theta_{12} \cos^2\theta_{13}e^{i\alpha_{21}} |,
\end{align}
where its observed value could be measured by KamLAND-Zen in future~\cite{KamLAND-Zen:2016pfg}. 
 $\kappa$ is given in terms of atmospheric mass square difference $\Delta m^2_{\rm atm}$ and dimensionless neutrino mass eigenvalues as follows:
\begin{align}
({\rm NH})&: \kappa^2_{\rm NH}=\frac{\Delta m^2_{\rm atm}}{\tilde D_{\nu_3}^2},\\
({\rm IH})&: \kappa^2_{\rm IH}=\frac{\Delta m^2_{\rm atm}}{\tilde D_{\nu_2}^2}.
\label{eq:k}
\end{align}
Then
 the solar neutrino mass square difference is also found as 
 \begin{align}
({\rm NH})&: \Delta m^2_{\rm sol}=\kappa^2_{\rm NH }\tilde D_{\nu_2}^2,\\
({\rm IH})&: \Delta m^2_{\rm sol}=\kappa^2_{\rm IH }(\tilde D_{\nu_2}^2 - \tilde D_{\nu_1}^2).
\end{align}

\section{Dark Matter}
We have two kinds of DM candidates; the lightest Majorana fermion $D_{N_1}$ or the lightest inert boson of $\eta_{R,I}$.
Here, we briefly discuss both the DM cases. Hereafter, we unify the symbol of DM mass as $m_\chi$.

\subsection{Fermionic DM}  
In case of a fermionic DM candidate; $D_{N_1}$, the main annihilation channels come from $y_\eta$ so that we can explain the relic density of DM.
The cross section, which can be expanded by relative velocity $v^{1/2}_{\rm rel}\approx0.3$, is p-wave dominant which is given by
\begin{align}
(\sigma v)_{\rm rel} \approx \frac{m_\chi^2 (m_\chi^4 + m_{\eta}^4)}{48 \pi  (m_\chi^2 + m_{\eta}^2)^4} v_{\rm rel}^2
\sum_{a,b=1}^3(|(Y^{\ell\dag}_\eta)_{1a} (Y^\ell_\eta)_{b1} |^2+|(Y^{\nu \dag}_\eta)_{1a} (Y^\nu_\eta)_{b1}|^2),
\end{align}
where we assume $m_{\eta^\pm}\approx m_{\eta_R}\approx m_{\eta_I} $ simply in order to satisfy the constraints of oblique parameters,
and $m_\ell/m_{\chi}\approx m_\ell/m_{\eta}\approx 0$.
Then, the relic density is approximately given by
\begin{align}
\Omega h^2\approx \frac{3.63\times 10^8 x_f^2}{3  g_*^{1/2} m_{\rm pl} (\sigma v)_{\rm rel}},
\end{align}
where $g_*\approx 100$ is the number of relativistic degrees of freedom when DM freezes out, and $m_{\rm pl}\approx1.22\times 10^{19}$ GeV  is the Planck mass.
The Planck experiment tells us~\cite{Planck:2013pxb}
\begin{align}
\Omega h^2 = 0.1196\pm 0.00031\ (68\%{\rm C.L.})\ .
\end{align}
The observed relic density corresponds to the following cross section within the 2 $\sigma$ interval:
\begin{align}
1.7756\times10^{-9} \lesssim  {(\sigma v)_{\rm rel}}\ {\rm GeV^{2}} \lesssim1.9697 \times10^{-9} .
\end{align}
In our numerical analysis, we apply the above value to find our solutions in case of fermionic DM.
Direct detection constraint is obtained by one-loop diagram through $y_\eta$ and we can easily evade this constraint~\cite{Abe:2018emu}.
Thus, we do not discuss here.

\subsection{Bosonic DM}  
DM is expected to be a real component of inert scalar $\eta$; $\eta_R$. In order to avoid a constraint of the oblique parameters, we assume to be {$m_{\eta^\pm} \approx m_I$} for simplicity.
In addition, we suppose the mass difference between $\eta_R$ and $\eta_I$ should be larger than ${\cal O}(100)$ keV in order to avoid the constraint of direct detection searches via Z-boson exchange. But we expect that the mass difference should be less than   ${\cal O}(1)$ GeV
to satisfy the oblique parameter's bounds.
In this case, the mass of DM is uniquely fixed by the observed relic density which suggests it is within $534\pm8.5$ GeV~\cite{Hambye:2009pw}~\footnote{In addition to this range, there exists a solution to satisfy the relic density and direct detection at nearby $m_\chi= m_h/2$ that is Higgs
portal  interactions. But, we do not discuss this solution because of its trivial solution.}, when the Yukawa coupling is not so large. In fact, tiny Yukawa couplings are requested by satisfying
the data. Thus, we just work on the mass of $\eta$ at this narrow range.
Note here that we cannot write a rather simple formula of the cross section to explain the relic density because we need to consider co-annihilation channels due to degeneracy among boson masses as briefly discussed above.
Fortunately, a preceding analysis in ref.~\cite{Hambye:2009pw} has included all our valid interactions to explain the relic density.
Thus, we just fit the DM mass to be in the range of $534\pm8.5$ GeV in our numerical analysis below.


\section{Numerical analysis}
In this section, we perform chi-square numerical analysis for the neutrino oscillation data
and demonstrate our predictions satisfying LFV bounds and DM constraints discussed above.
Here, we apply Nufit 6.0 for the neutrino oscillation data~\cite{Esteban:2024eli}.
Note that Dirac CP phase $\delta_{\rm CP}$ is not considered as the observed value but done as the predicted one.
Thus, we consider only five observed values; three mixings and two mass squared differences.
Our input parameters are randomly selected in the following ranges:
\begin{align}
(|\tilde M_1|,\tilde m_R) = [10^{-2}, 10^2]\ {\rm GeV}, \
(\tilde \alpha_\nu,\tilde \beta_\nu) = [10^{-3}, 10^3], \
{\rm Arg}[\tilde M_1] = [-\pi, \pi],\\
\gamma_\nu = [10^{-10}, 1],
\end{align}
where we work $\tau$ on the fundamental region, and $\tilde m_I = \tilde m_{\eta^\pm}$.

\subsection{NH}
Here, we show several predictions in the case of NH hierarchy including features of DM.

\subsubsection{Fermionic DM}

\begin{figure}[tb]\begin{center}
\includegraphics[width=53mm]{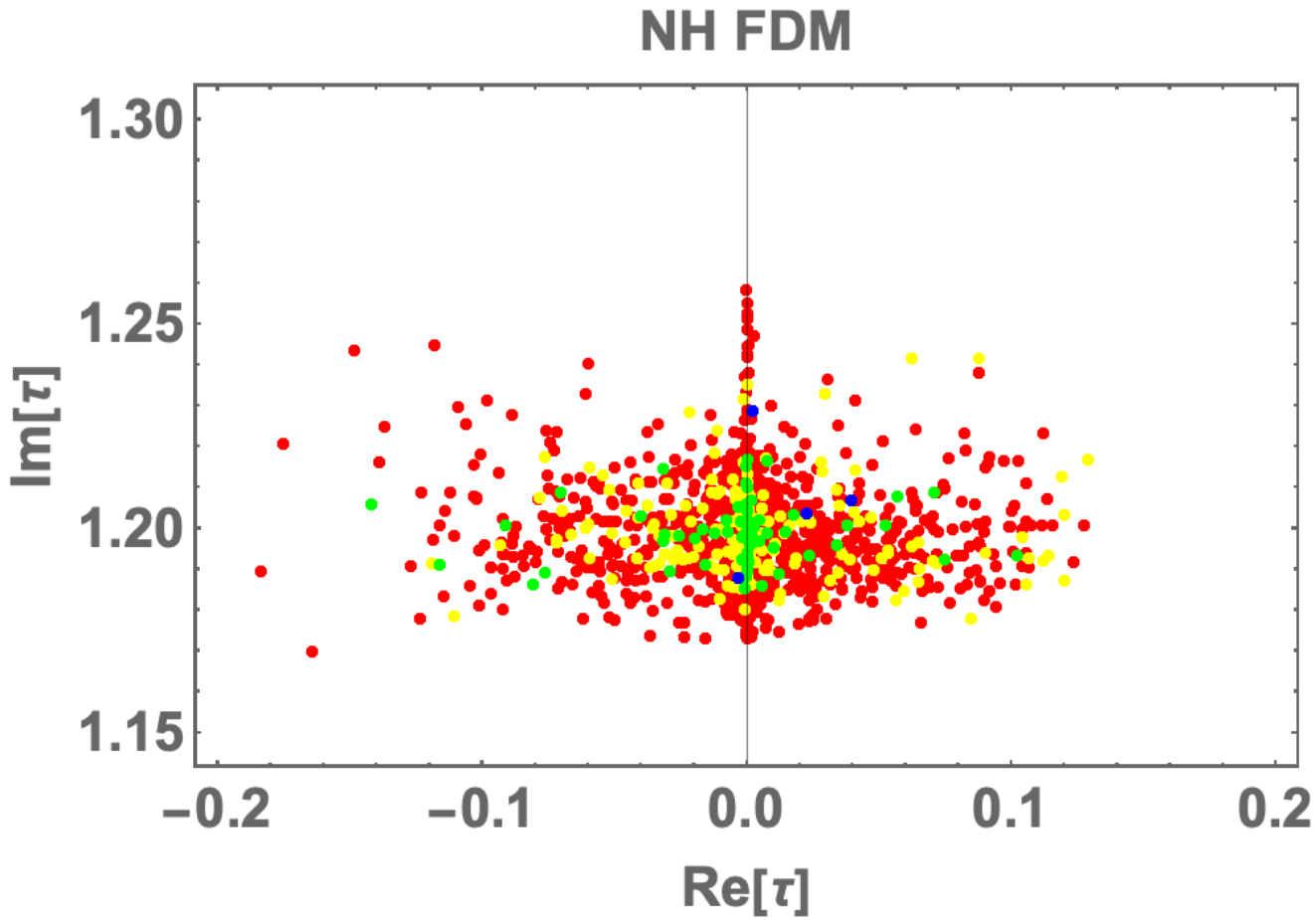}
\includegraphics[width=53mm]{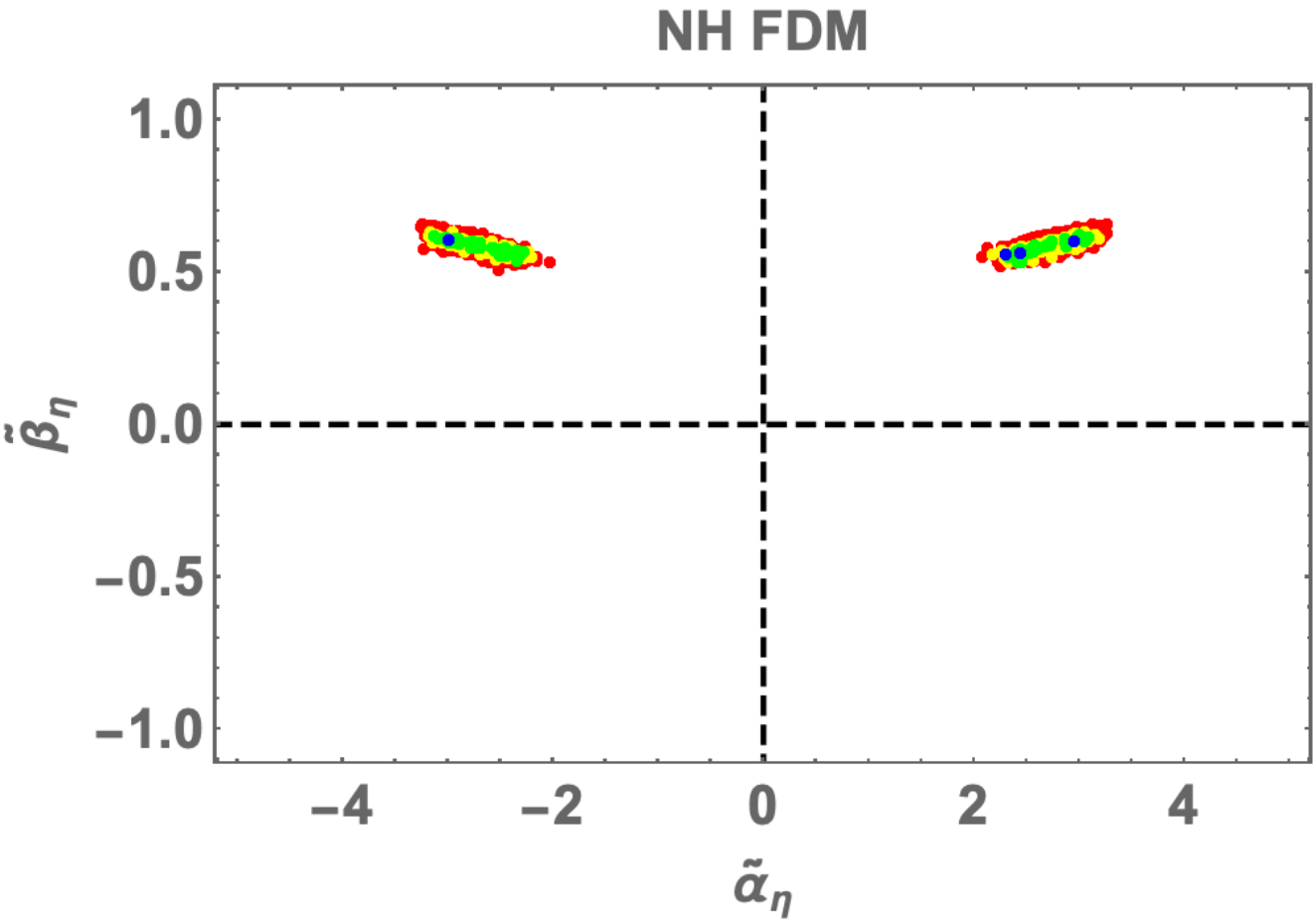}
\includegraphics[width=53mm]{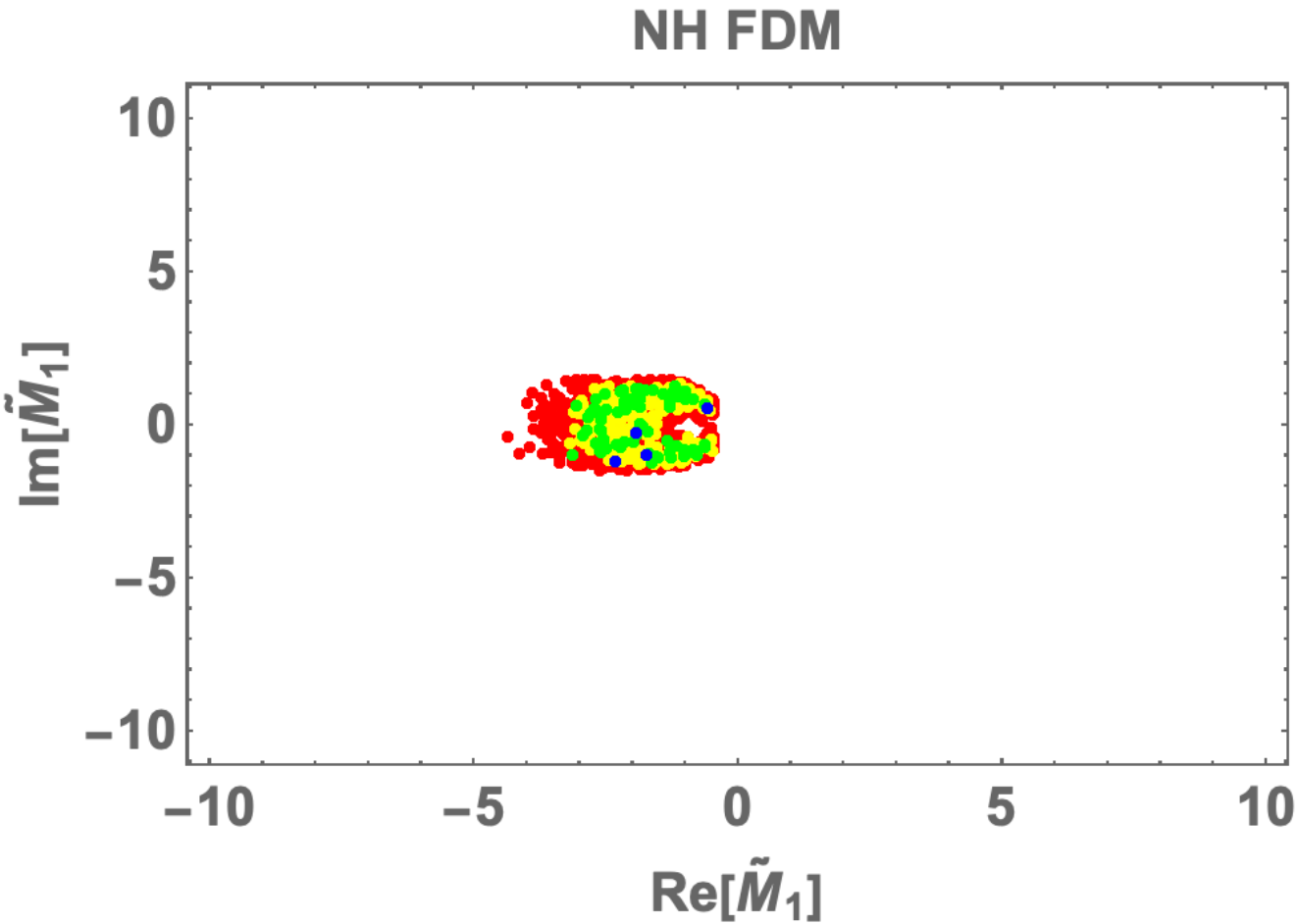}
\caption{Allowed regions for input parameters $\tau$(left-side), $\tilde\alpha$ and $\tilde\beta$(center), and $\tilde M_1$(right-side) where the blue, green, yellow, and red points respectively represent the intervals of 
$<1\sigma$, $1\sigma-2\sigma$,
$2\sigma-3\sigma$, and $3\sigma-5\sigma$. } 
\label{fig:nhfdm1}
\end{center}\end{figure}

In Fig.~\ref{fig:nhfdm1}, we show the allowed ranges for our input parameters; $\tau$(left-side), $\tilde\alpha$ and $\tilde\beta$(center), and $\tilde M_1$(right-side) where the blue, green, yellow, and red points respectively represent the intervals of 
$<1\sigma$, $1\sigma-2\sigma$,
$2\sigma-3\sigma$, and $3\sigma-5\sigma$.
The left-side figure tells us $-0.2\lesssim {\rm Re}[\tau]\lesssim 0.14$ and $1.17\lesssim {\rm Im}[\tau]\lesssim 1.26$.
The center-side figure suggests $2\lesssim |\tilde\alpha|\lesssim 3.5$ and $0.5\lesssim \tilde\beta \lesssim 0.7$.
The right-side figure implies $-5\lesssim {\rm Re}[\tilde M_1] \lesssim 0$ and $-1.5\lesssim {\rm Im}[\tilde M_1] \lesssim 1.5$.

\begin{figure}[tb]\begin{center}
\includegraphics[width=80mm]{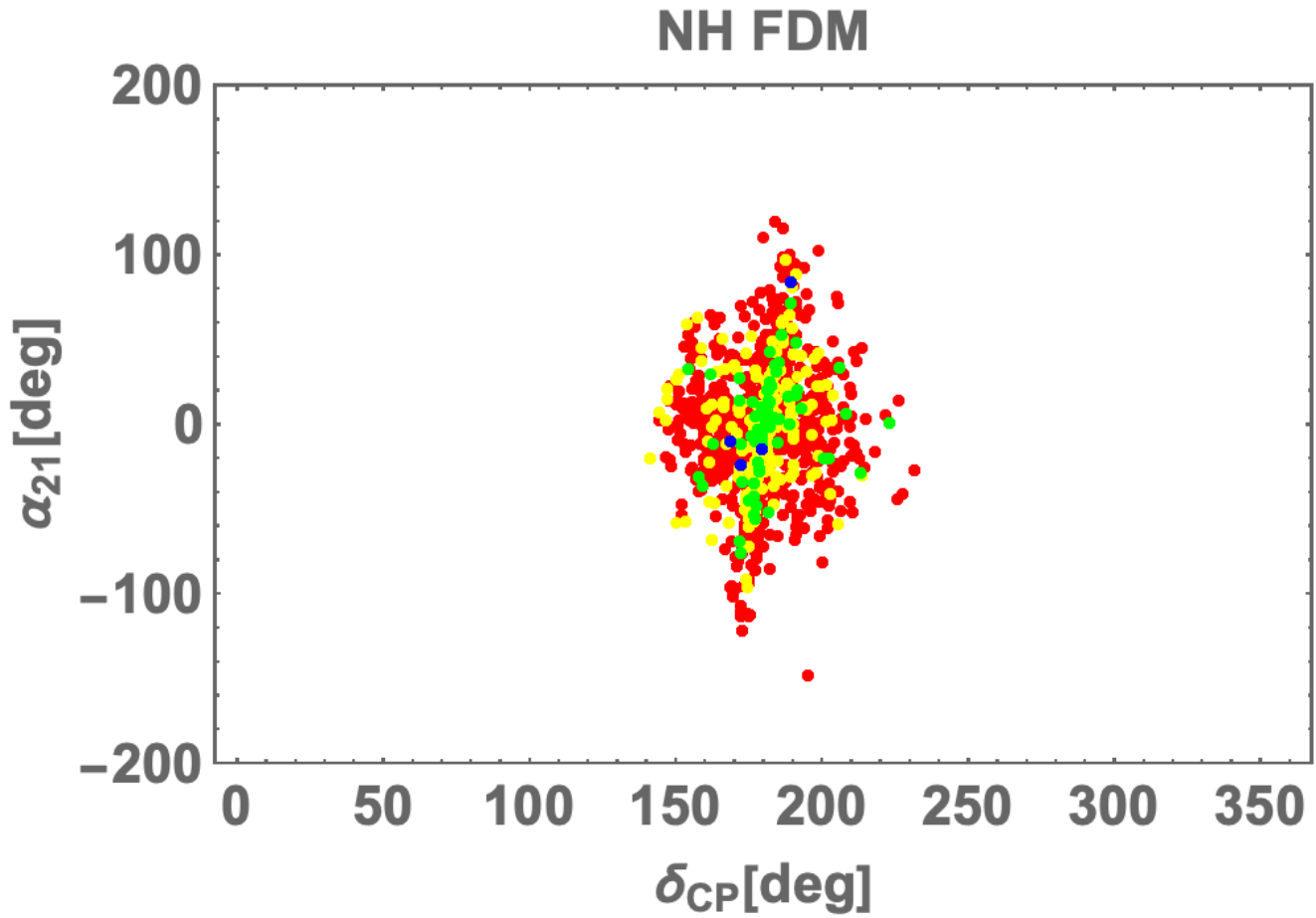}
\includegraphics[width=80mm]{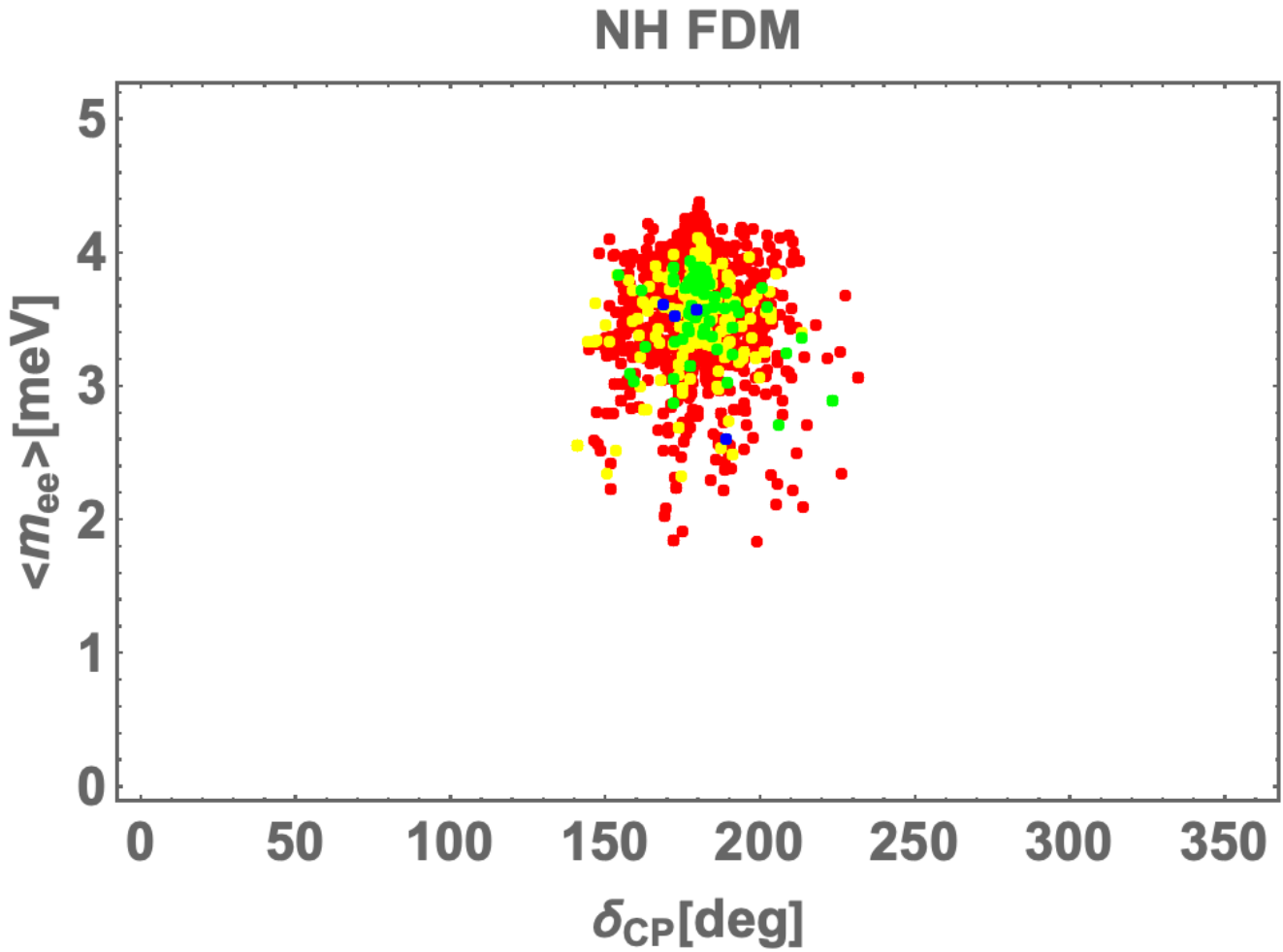}\\
\includegraphics[width=80mm]{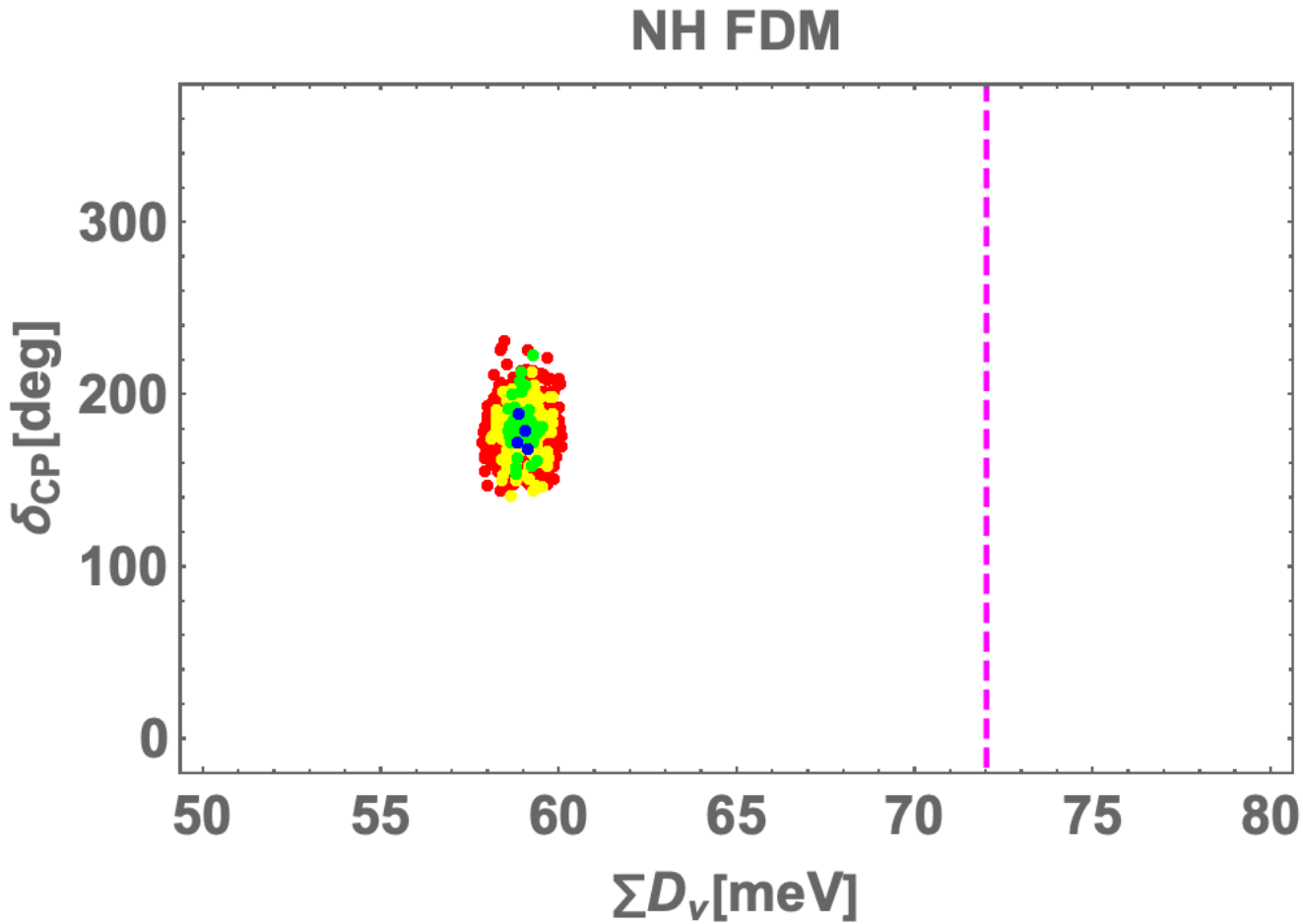}
\includegraphics[width=80mm]{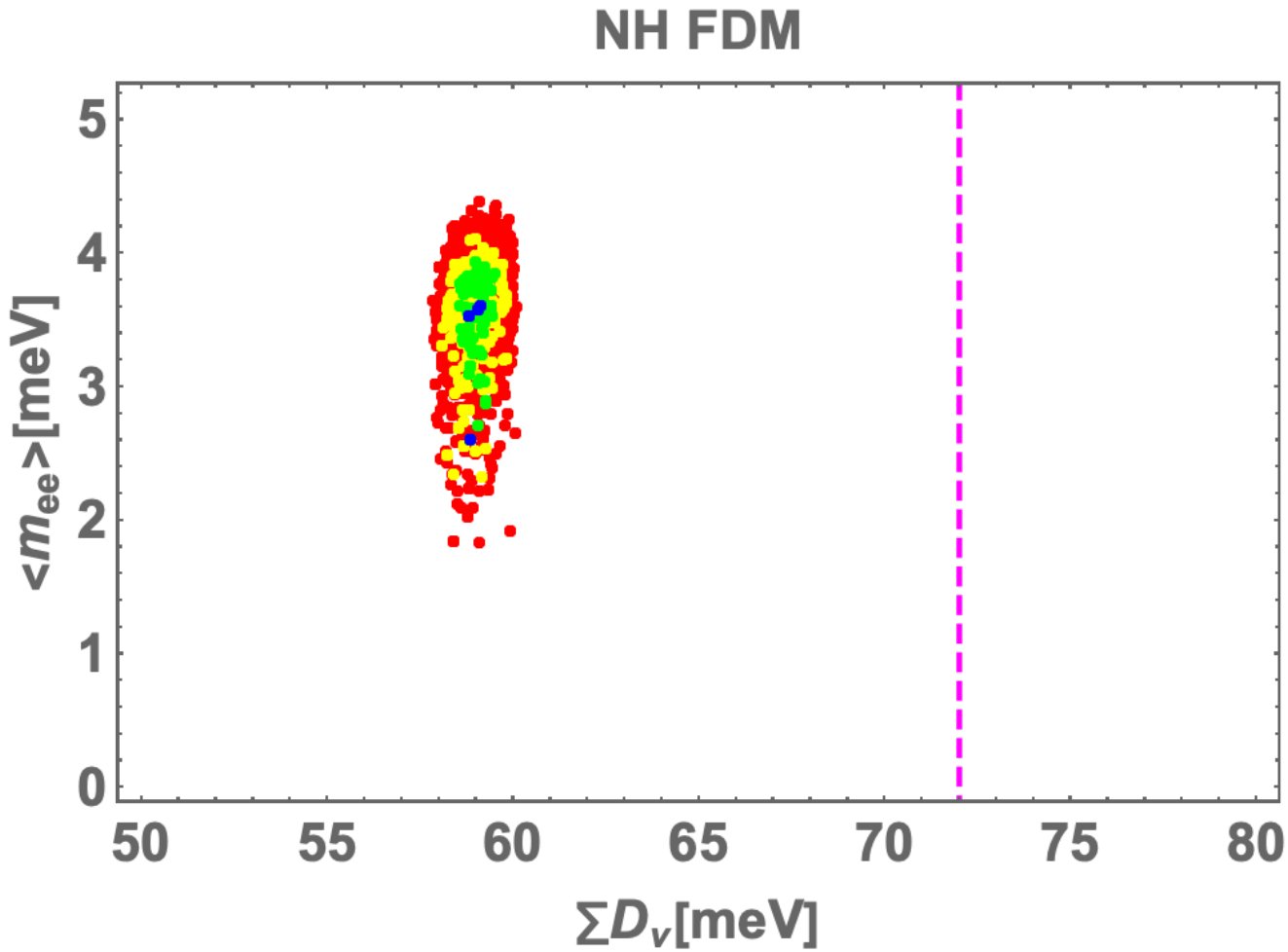}
\caption{Allowed regions for Majorana phase $\alpha_{21}$ in terms of $\delta_{\rm CP}$(up-left), $\langle m_{ee}\rangle$ in terms of $\delta_{\rm CP}$ (up-right), $\delta_{\rm CP}$ in terms of $\sum D_\nu$ (down-left), and  $\langle m_{ee}\rangle$ in terms of $\sum D_\nu$ (down-right). The color legends are the same as Fig.~\ref{fig:nhfdm1}. In down figures, the vertical dotted magenta lines are the experimental result  DESI and CMB data combination.}   
\label{fig:nhfdm2}\end{center}\end{figure}

Fig.~\ref{fig:nhfdm2} shows allowed regions for Majorana phase $\alpha_{21}$ in terms of $\delta_{\rm CP}$(up-left), $\langle m_{ee}\rangle$ in terms of $\delta_{\rm CP}$ (up-right), $\delta_{\rm CP}$ in terms of $\sum D_\nu$ (down-left), and  $\langle m_{ee}\rangle$ in terms of $\sum D_\nu$ (down-right). The color legends are the same as Fig.~\ref{fig:nhfdm1}. In down figures, the vertical dotted magenta lines are the experimental result  DESI and CMB data combination $\sum D_\nu\le$ 72 meV.
These figures predict to be $140^\circ\lesssim \delta_{\rm CP}\lesssim 240^\circ$ and  $0^\circ\lesssim \alpha_{21}\lesssim 120^\circ,\
210^\circ\lesssim \alpha_{21}\lesssim 360^\circ$, $2\ {\rm meV}\lesssim\langle m_{ee}\rangle\lesssim 4.4\ {\rm meV}$,
and $57\ {\rm meV}\lesssim\sum D_\nu\lesssim60\ {\rm meV}$.
All figures are rather localized at narrow regions but there is not so strong or specific correlations.

\begin{figure}[tb]\begin{center}
\includegraphics[width=80mm]{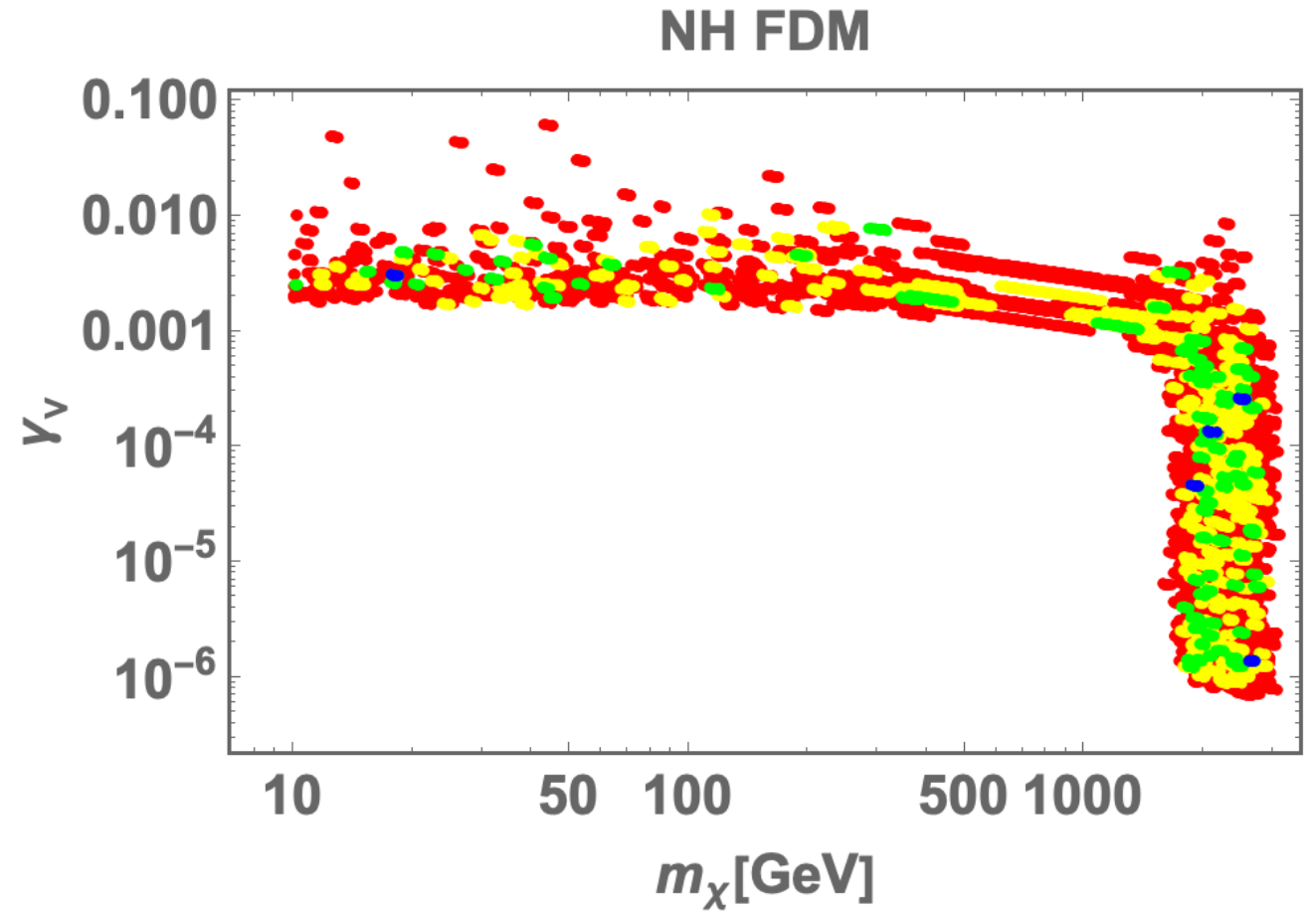}
\caption{Allowed region of $\gamma_\nu$ in terms of DM mass where the color legends are the same as Fig.~\ref{fig:nhfdm1}.}   
\label{fig:nhfdm3}\end{center}\end{figure}
Fig.~\ref{fig:nhfdm3} demonstrates allowed region of $\gamma_{\nu}$ in terms of $m_{\chi}$ where the color legends are the same as Fig.~\ref{fig:nhfdm1}.
The figure suggests that our DM mass range is [10, 3600] GeV and $\gamma_\nu$ is [$10^{-6}$, 0.1] with rather strong correlation.

In LFVs, we have obtained their maximum branching ratios $4.20\times10^{-13}$ for $\mu\to e\gamma$,
 $1.40\times10^{-10}$ for $\tau\to e\gamma$, and  $4.66\times10^{-9}$ for $\tau\to \mu\gamma$.
Thus, $\mu\to e\gamma$ would be the most verifiable branching ratio in the experiments.
\footnote{Moreover, its future experiment MEG II suggests the branching ratio would reach at $6\times 10^{-14}$~\cite{Venturini:2024keu}.}
In a future experiment, Super B factory~\cite{Aushev:2010bq} might reach at $10^{-9}$ for $\tau\to \mu\gamma$.
Hence, $\tau\to \mu\gamma$ would be also a promising mode to be detected by the future experiment.

\subsubsection{Bosonic DM}

\begin{figure}[tb]\begin{center}
\includegraphics[width=53mm]{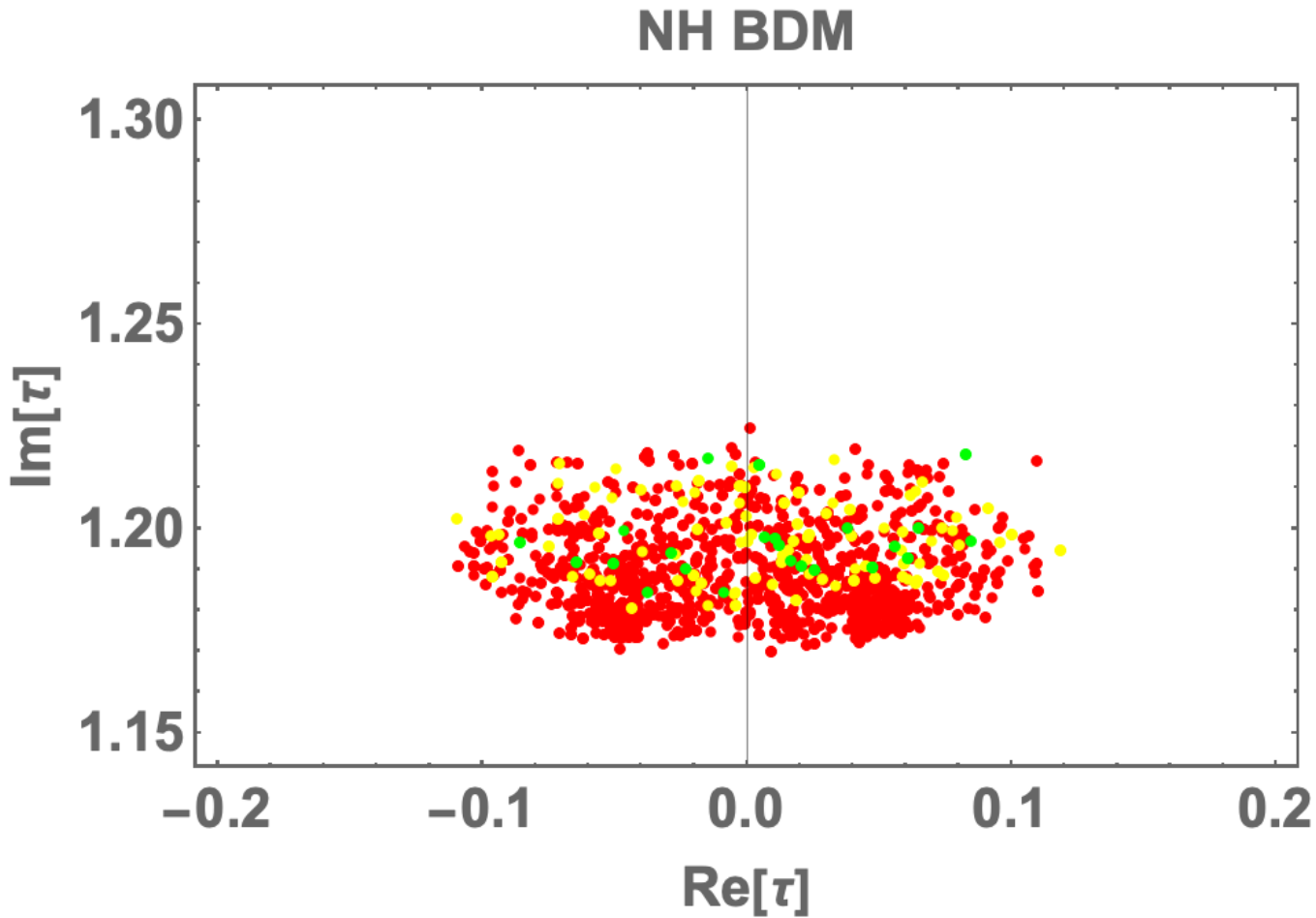}
\includegraphics[width=53mm]{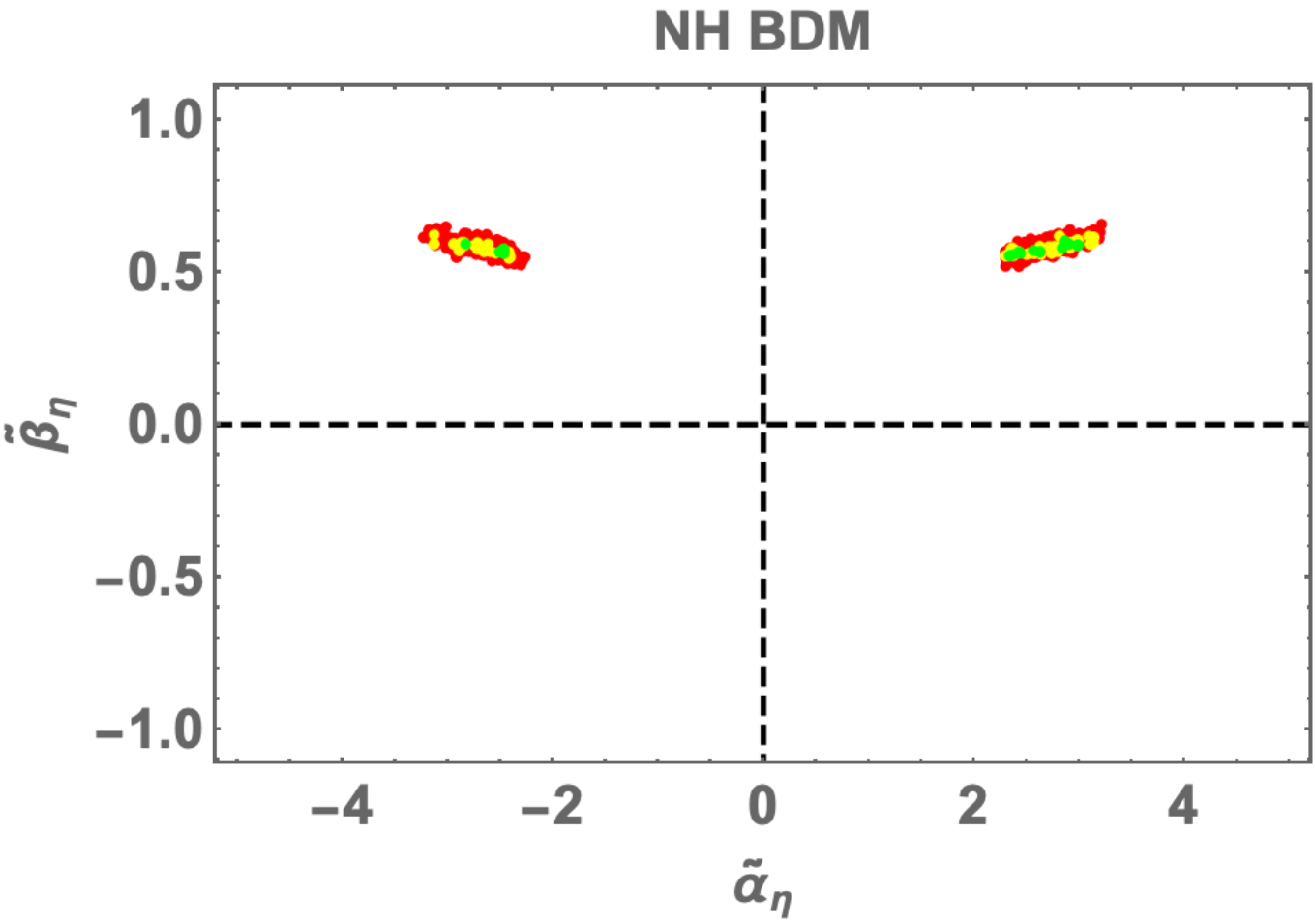}
\includegraphics[width=53mm]{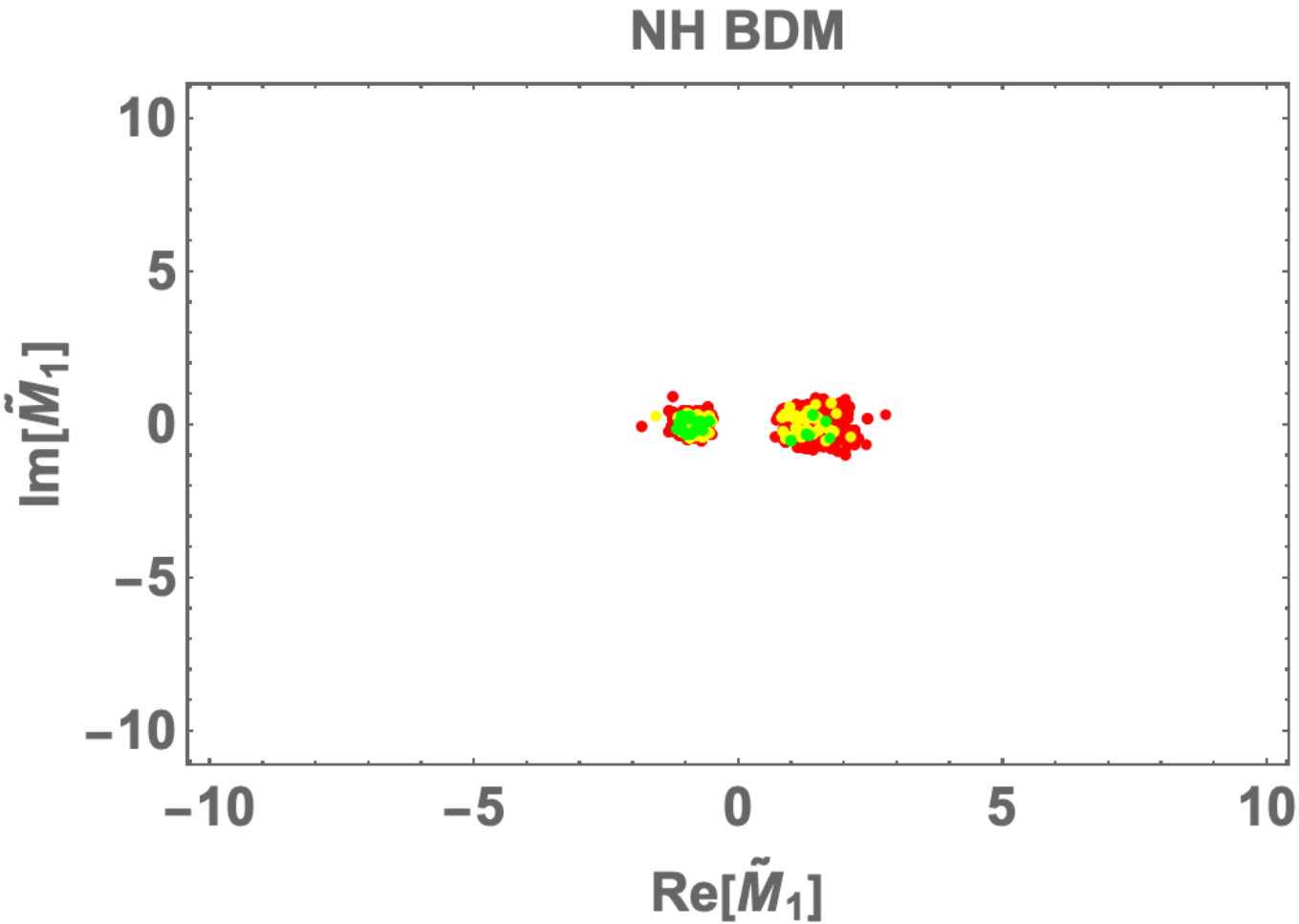}
\caption{Allowed regions for input parameters $\tau$(left-side), $\tilde\alpha$ and $\tilde\beta$(center), and $\tilde M_1$(right-side), where the color legends are the same as Fig.~\ref{fig:nhfdm1}. } 
\label{fig:nhbdm1}
\end{center}\end{figure}

In Fig.~\ref{fig:nhbdm1}, we show the allowed ranges for our input parameters; $\tau$(left-side), $\tilde\alpha$ and $\tilde\beta$(center), and $\tilde M_1$(right-side) where the color legends are the same as Fig.~\ref{fig:nhfdm1}.
The left-side figure tells us $-0.1\lesssim {\rm Re}[\tau]\lesssim 0.1$ and $1.165\lesssim {\rm Im}[\tau]\lesssim 1.22$.
The center-side figure suggests $2\lesssim |\tilde\alpha|\lesssim 3.5$ and $0.5\lesssim \tilde\beta \lesssim 0.65$ which are almost the same region as the case of NH.
The right-side figure implies $-2\lesssim {\rm Re}[\tilde M_1] \lesssim -0.5, \ 0.5\lesssim {\rm Re}[\tilde M_1] \lesssim 2.5$ and $0\lesssim |{\rm Im}[\tilde M_1]| \lesssim 1$.

\begin{figure}[tb]\begin{center}
\includegraphics[width=80mm]{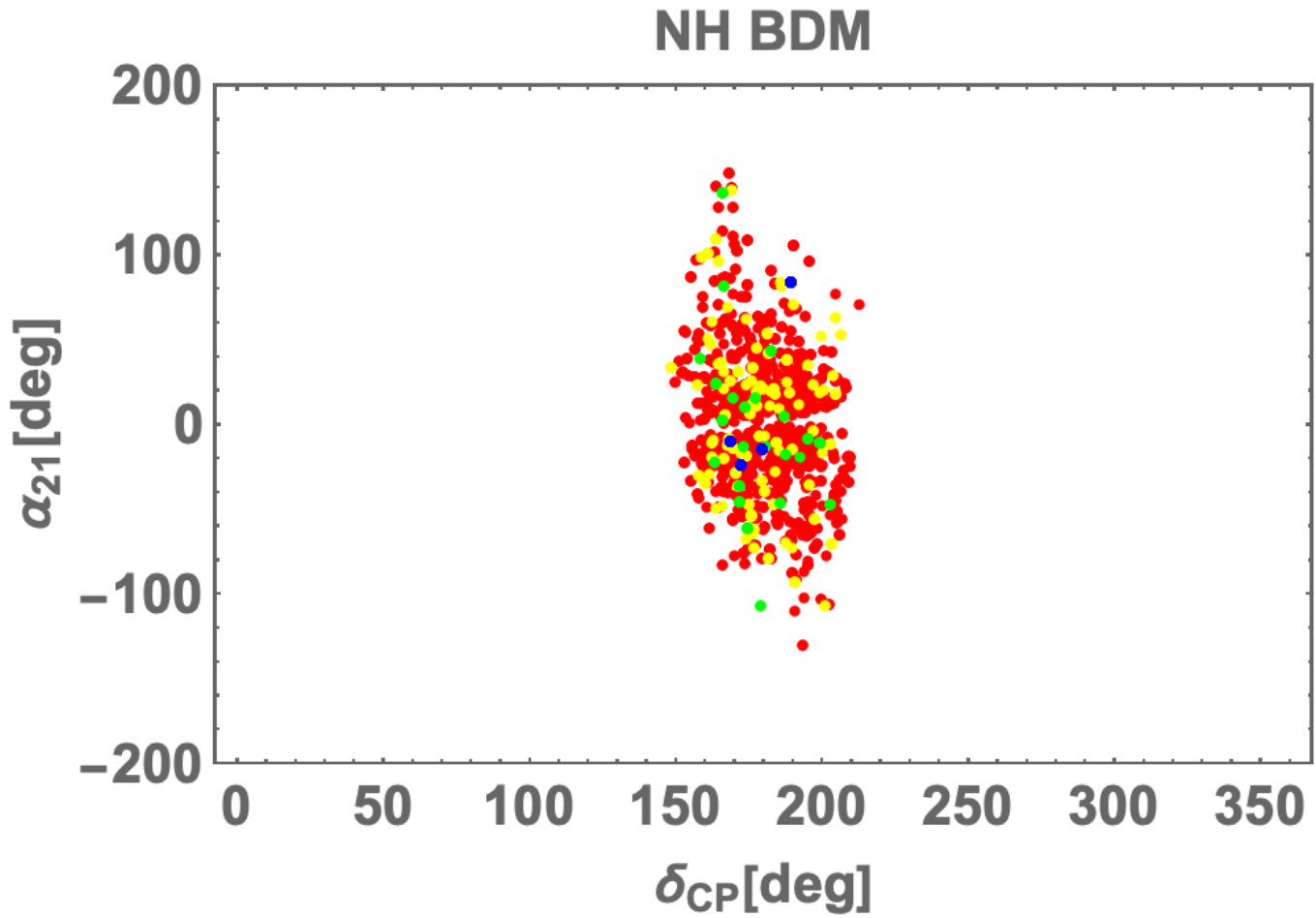}
\includegraphics[width=80mm]{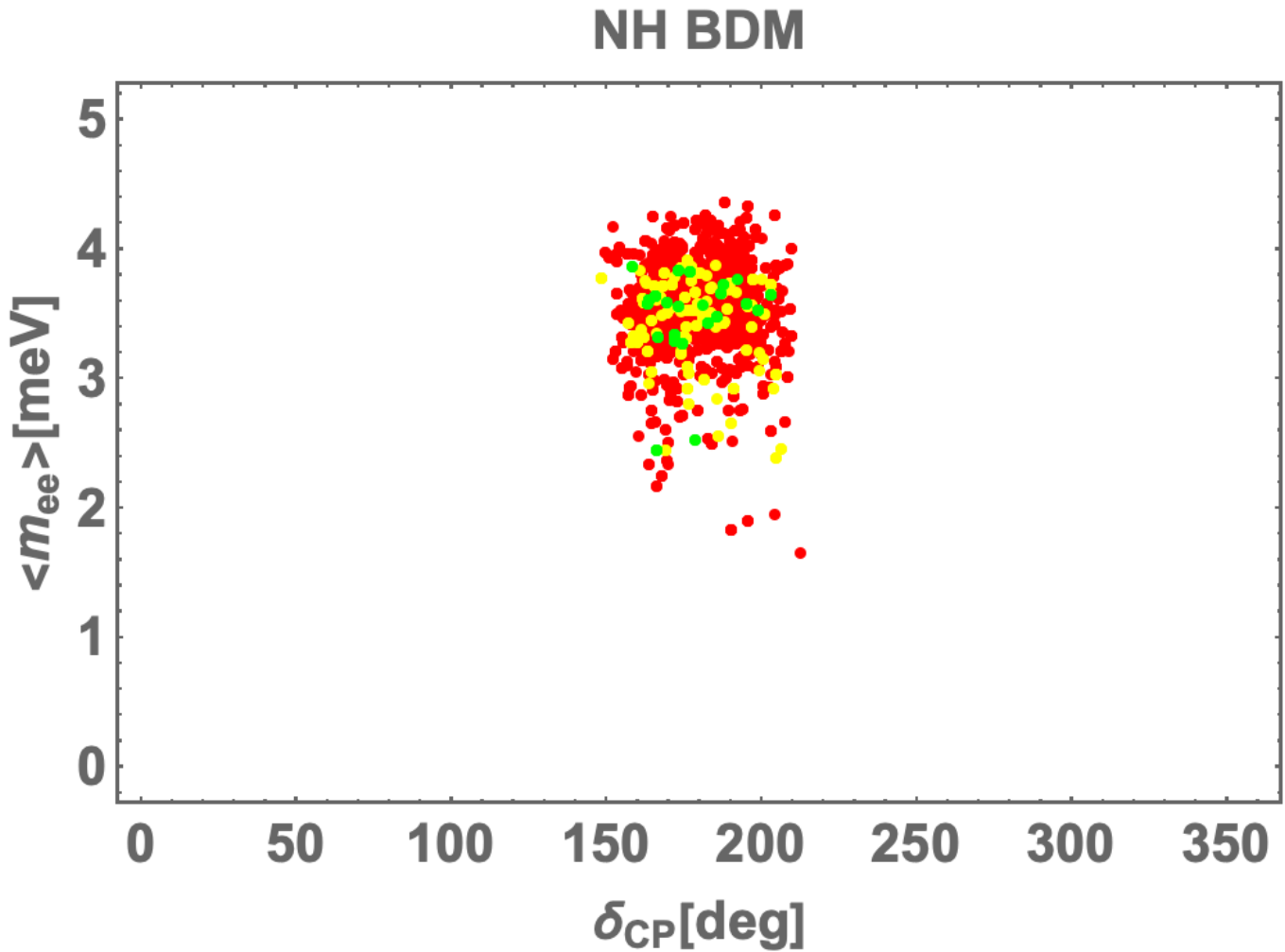}\\
\includegraphics[width=80mm]{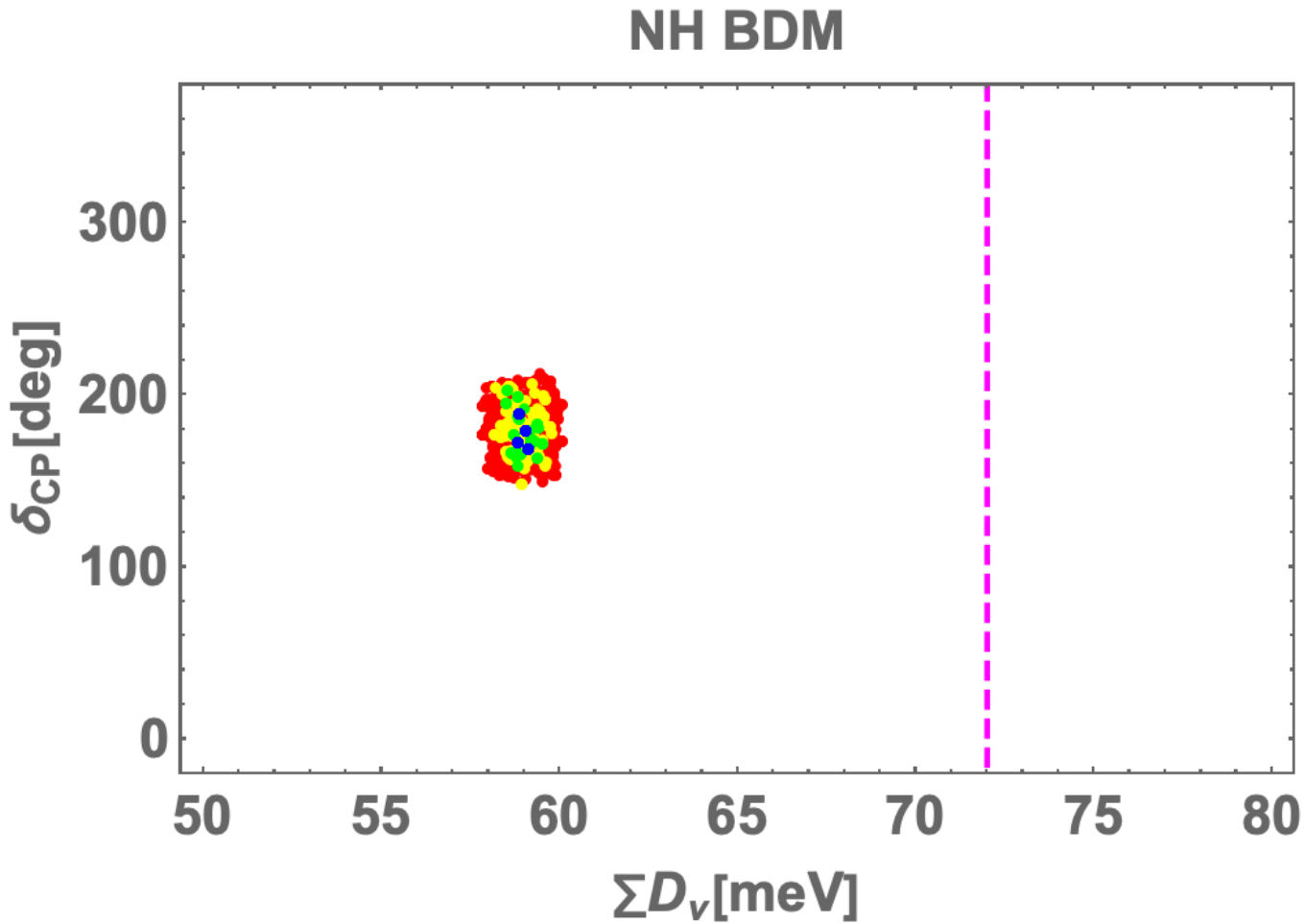}
\includegraphics[width=80mm]{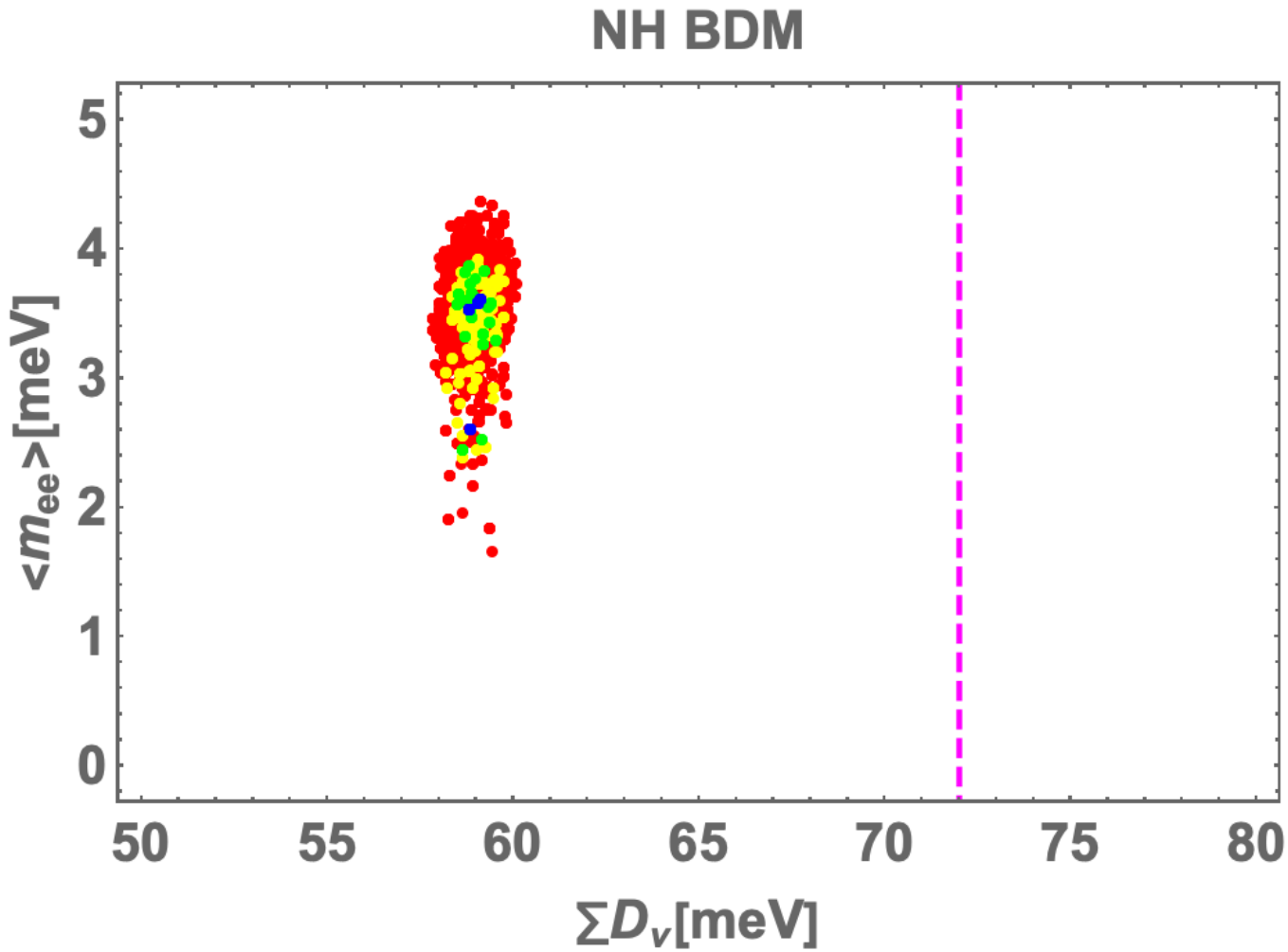}
\caption{Allowed regions for Majorana phase $\alpha_{21}$ in terms of $\delta_{\rm CP}$(up-left), $\langle m_{ee}\rangle$ in terms of $\delta_{\rm CP}$ (up-right), $\delta_{\rm CP}$ in terms of $\sum D_\nu$ (down-left), and  $\langle m_{ee}\rangle$ in terms of $\sum D_\nu$ (down-right). The color legends are the same as Fig.~\ref{fig:nhfdm1}. In down figures, the vertical dotted magenta lines are the experimental result  DESI and CMB data combination.}   
\label{fig:nhbdm2}\end{center}\end{figure}

Fig.~\ref{fig:nhbdm2} shows allowed regions for Majorana phase $\alpha_{21}$ in terms of $\delta_{\rm CP}$(up-left), $\langle m_{ee}\rangle$ in terms of $\delta_{\rm CP}$ (up-right), $\delta_{\rm CP}$ in terms of $\sum D_\nu$ (down-left), and  $\langle m_{ee}\rangle$ in terms of $\sum D_\nu$ (down-right). The color legends are the same as Fig.~\ref{fig:nhfdm1}. 
These figures predict to be $150^\circ\lesssim \delta_{\rm CP}\lesssim 210^\circ$ and  $0^\circ\lesssim \alpha_{21}\lesssim 150^\circ,\
230^\circ\lesssim \alpha_{21}\lesssim 360^\circ$, $1.8\ {\rm meV} \lesssim\langle m_{ee}\rangle\lesssim 4.4\ {\rm meV}$,
and $58\ {\rm meV}\lesssim\sum D_\nu\lesssim 60\ {\rm meV}$.
Tendencies of all figures are similar to the case of NH.

\begin{figure}[tb]\begin{center}
\includegraphics[width=80mm]{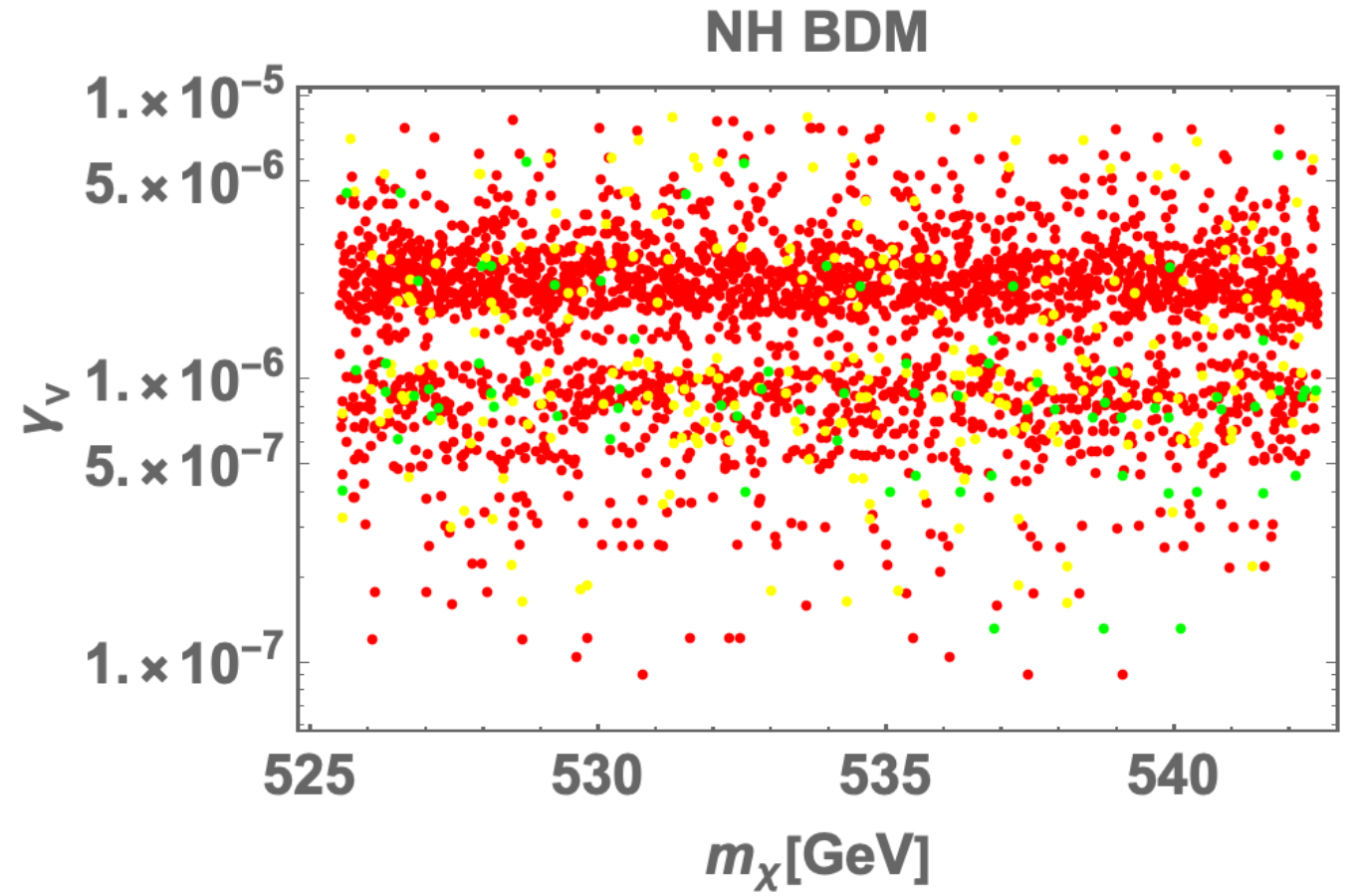}
\caption{Allowed region of $\gamma_\nu$ in terms of DM mass where the color legends are the same as Fig.~\ref{fig:nhfdm1}.}   
\label{fig:nhbdm3}\end{center}\end{figure}
Fig.~\ref{fig:nhbdm3} demonstrates allowed region of $\gamma_{\nu}$ in terms of $m_{\chi}$ where the color legends are the same as Fig.~\ref{fig:nhfdm1}.
The figure suggests $\gamma_\nu$ has to be small; $10^{-7}-10^{-5}$, due to satisfying the constraints of LFVs.

In LFVs, we have obtained their maximum branching ratios $4.30\times10^{-22}$ for $\mu\to e\gamma$,
 $1.07\times10^{-22}$ for $\tau\to e\gamma$, and  $1.71\times10^{-20}$ for $\tau\to \mu\gamma$.
Thus, these branching ratios are far from the experimental bounds in Eq.~(\ref{eq:lfvs-cond}).

\subsection{IH}
Here, we show several predictions in case of IH hierarchy including feature of DM.

\subsubsection{Fermionic DM}

\begin{figure}[tb]\begin{center}
\includegraphics[width=53mm]{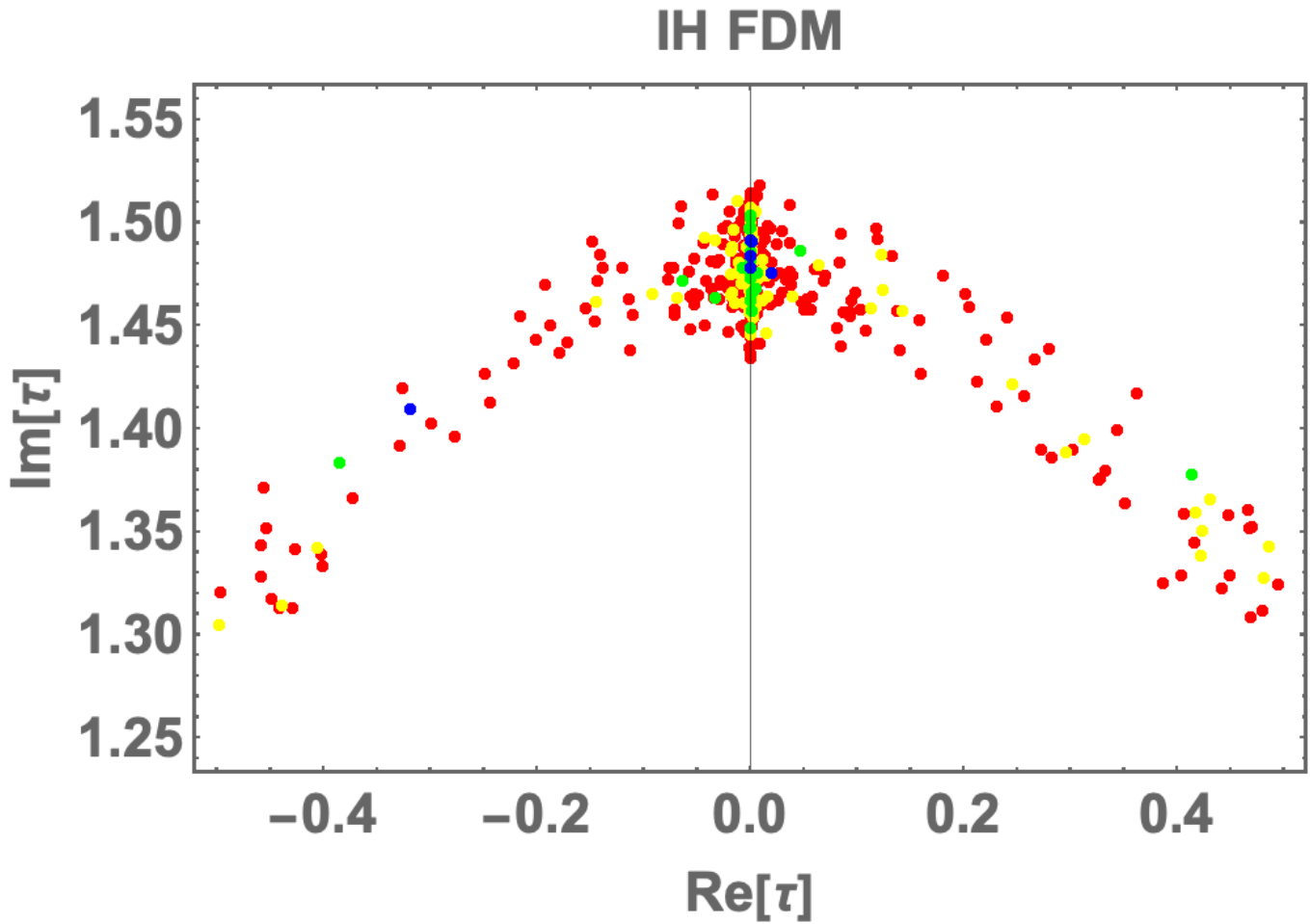}
\includegraphics[width=53mm]{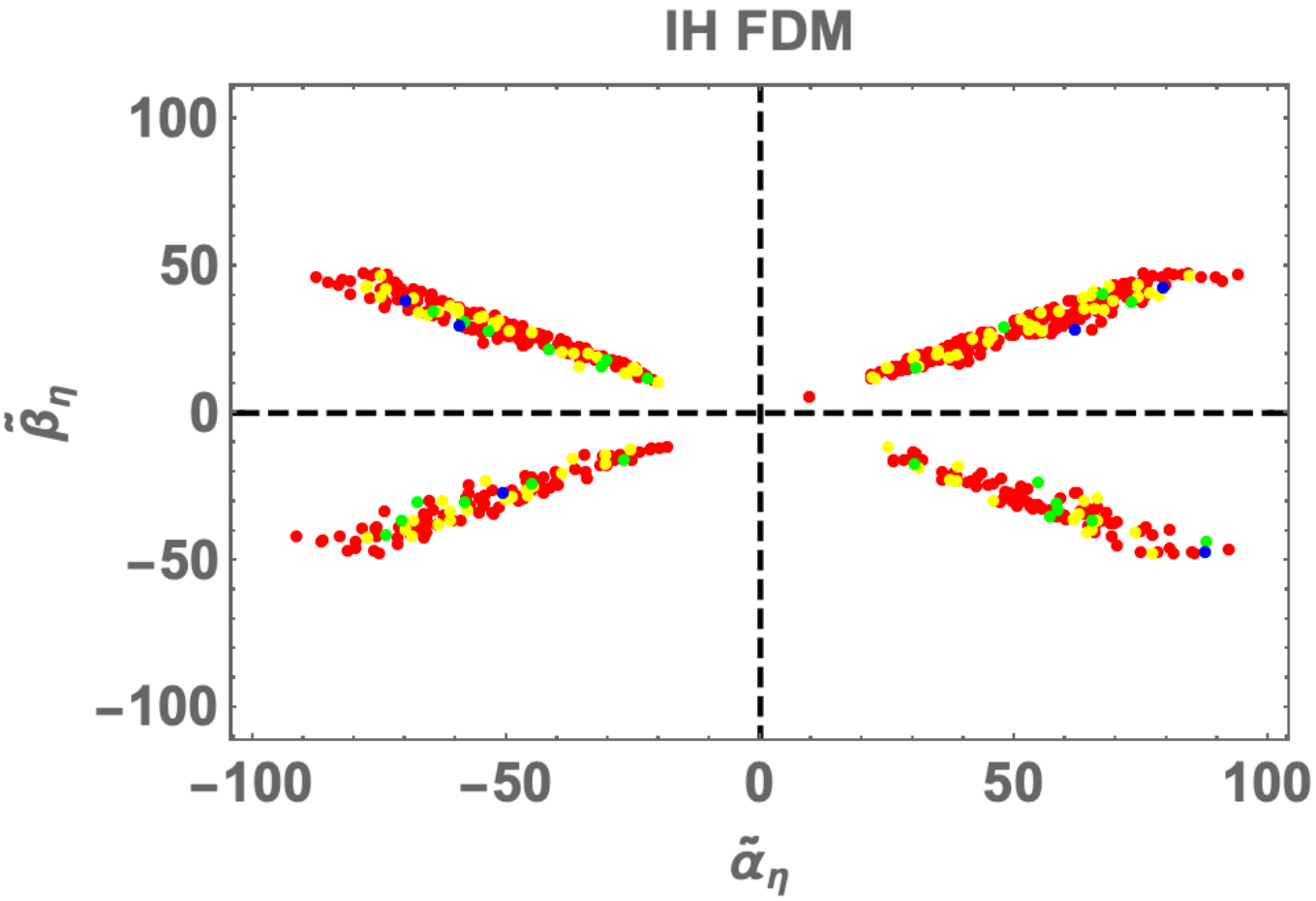}
\includegraphics[width=53mm]{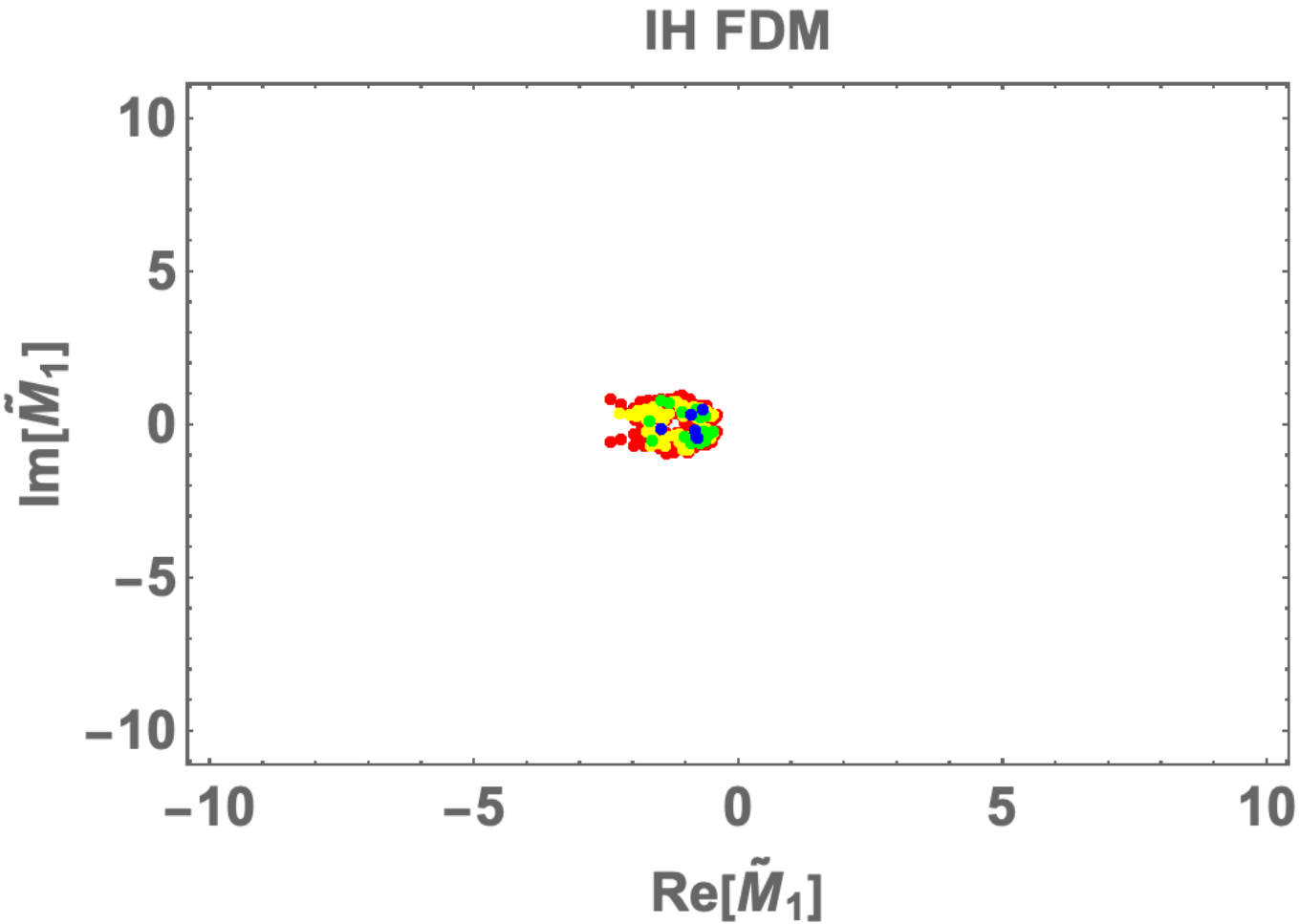}
\caption{Allowed regions for input parameters $\tau$(left-side), $\tilde\alpha$ and $\tilde\beta$(center), and $\tilde M_1$(right-side)  where the color legends are the same as Fig.~\ref{fig:nhfdm1}. } 
\label{fig:ihfdm1}
\end{center}\end{figure}

In Fig.~\ref{fig:ihfdm1}, we show the allowed ranges for our input parameters; $\tau$(left-side), $\tilde\alpha$ and $\tilde\beta$(center), and $\tilde M_1$(right-side)  where the color legends are the same as Fig.~\ref{fig:nhfdm1}.
The left-side figure tells us $ |{\rm Re}[\tau]|\lesssim 0.5$ and $1.30\lesssim {\rm Im}[\tau]\lesssim 1.50$.
The center-side figure suggests $10\lesssim |\tilde\alpha|\lesssim 100$ and $1\lesssim |\tilde\beta| \lesssim 50$.
The right-side figure implies $-3\lesssim {\rm Re}[\tilde M_1] \lesssim 0$ and $-1.5\lesssim {\rm Im}[\tilde M_1] \lesssim 1.5$.

\begin{figure}[tb]\begin{center}
\includegraphics[width=80mm]{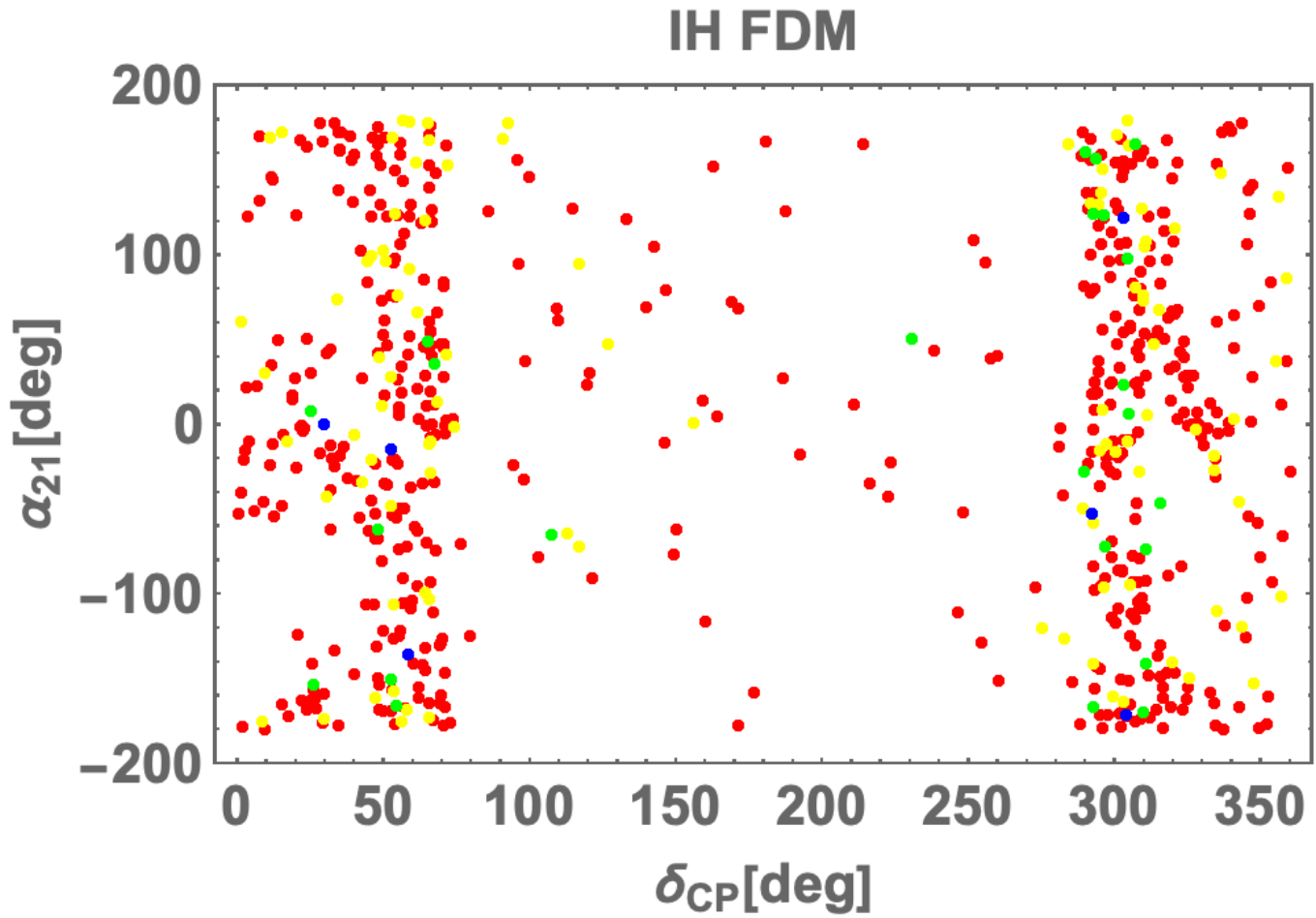}
\includegraphics[width=80mm]{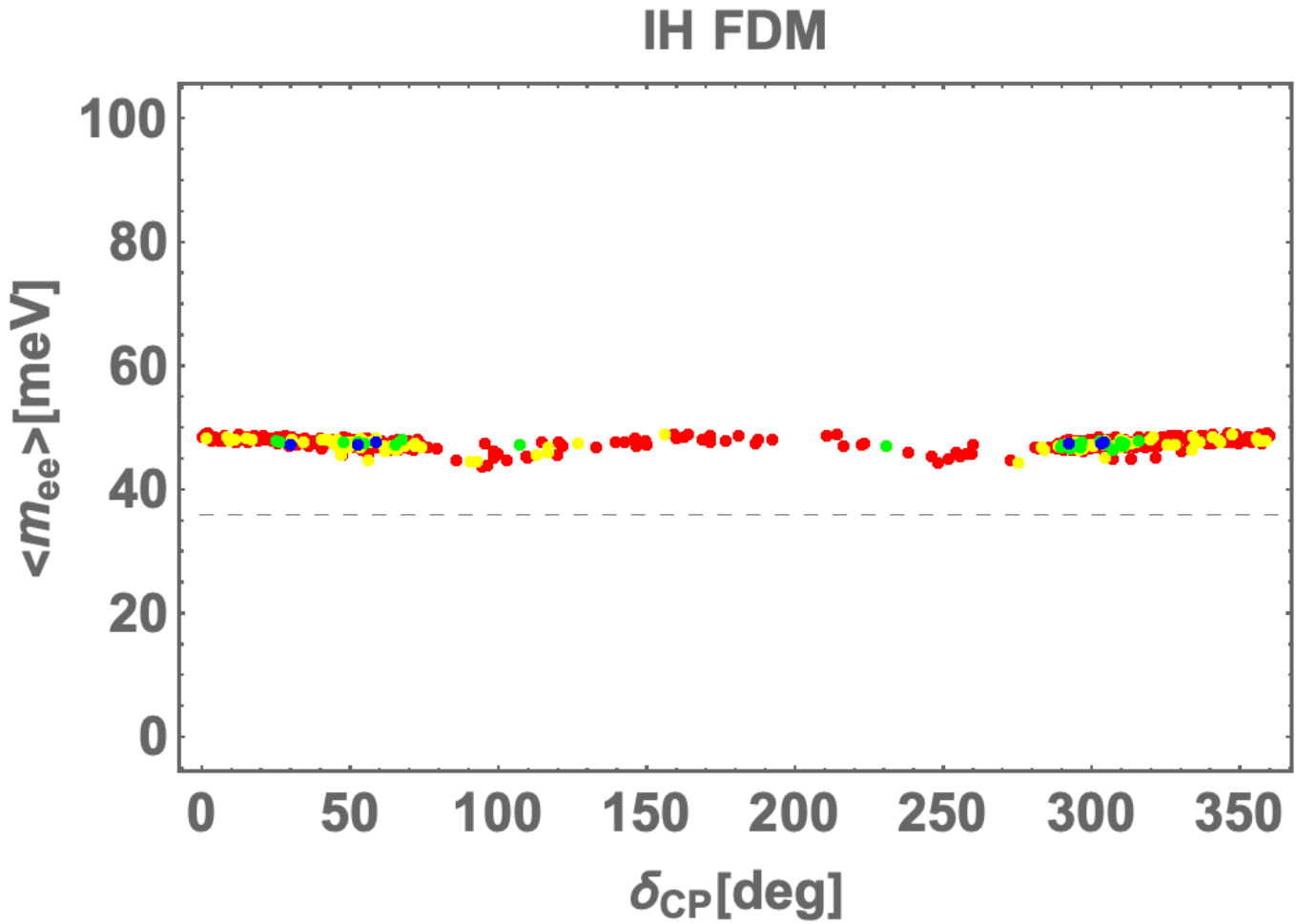}\\
\includegraphics[width=80mm]{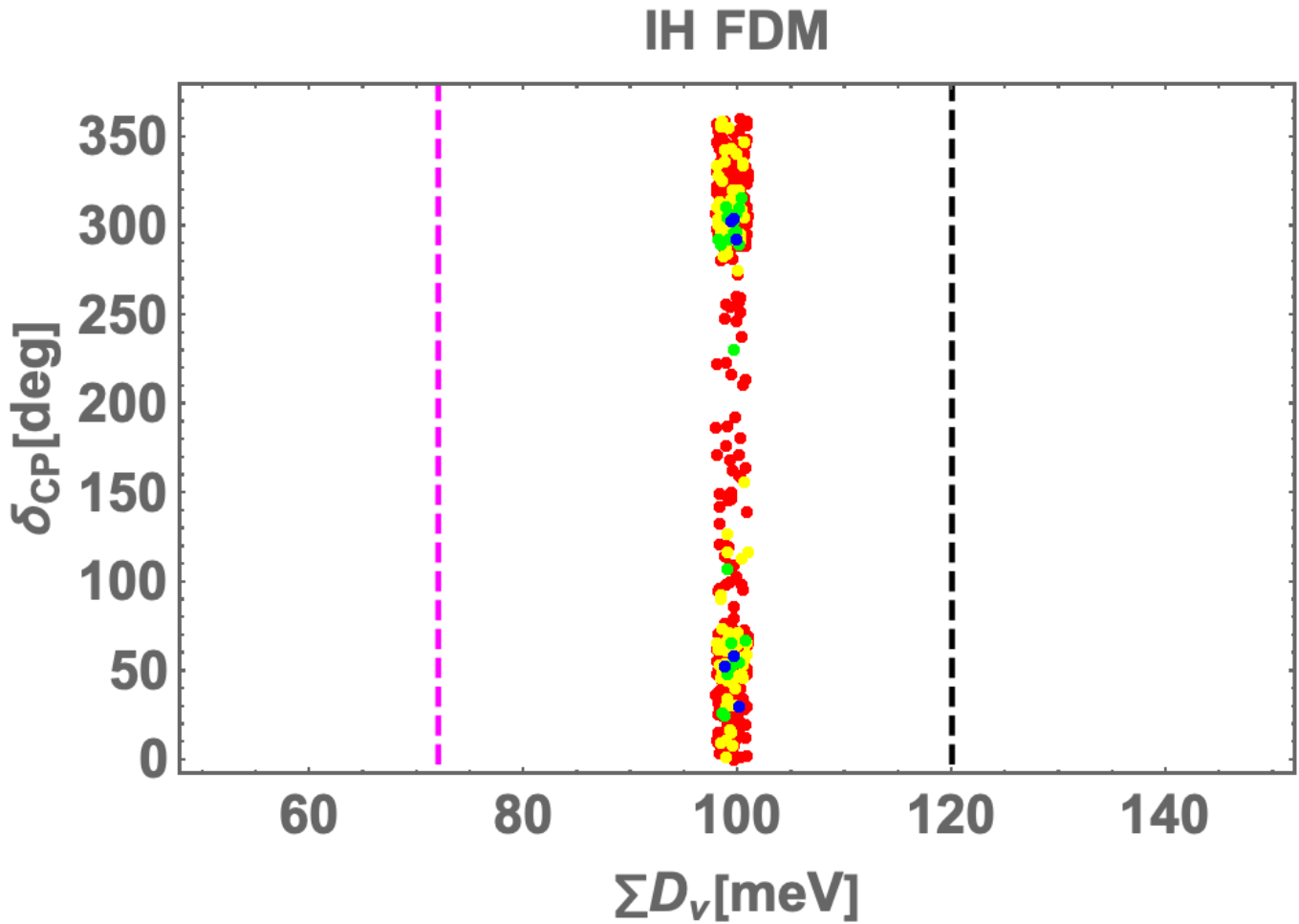}
\includegraphics[width=80mm]{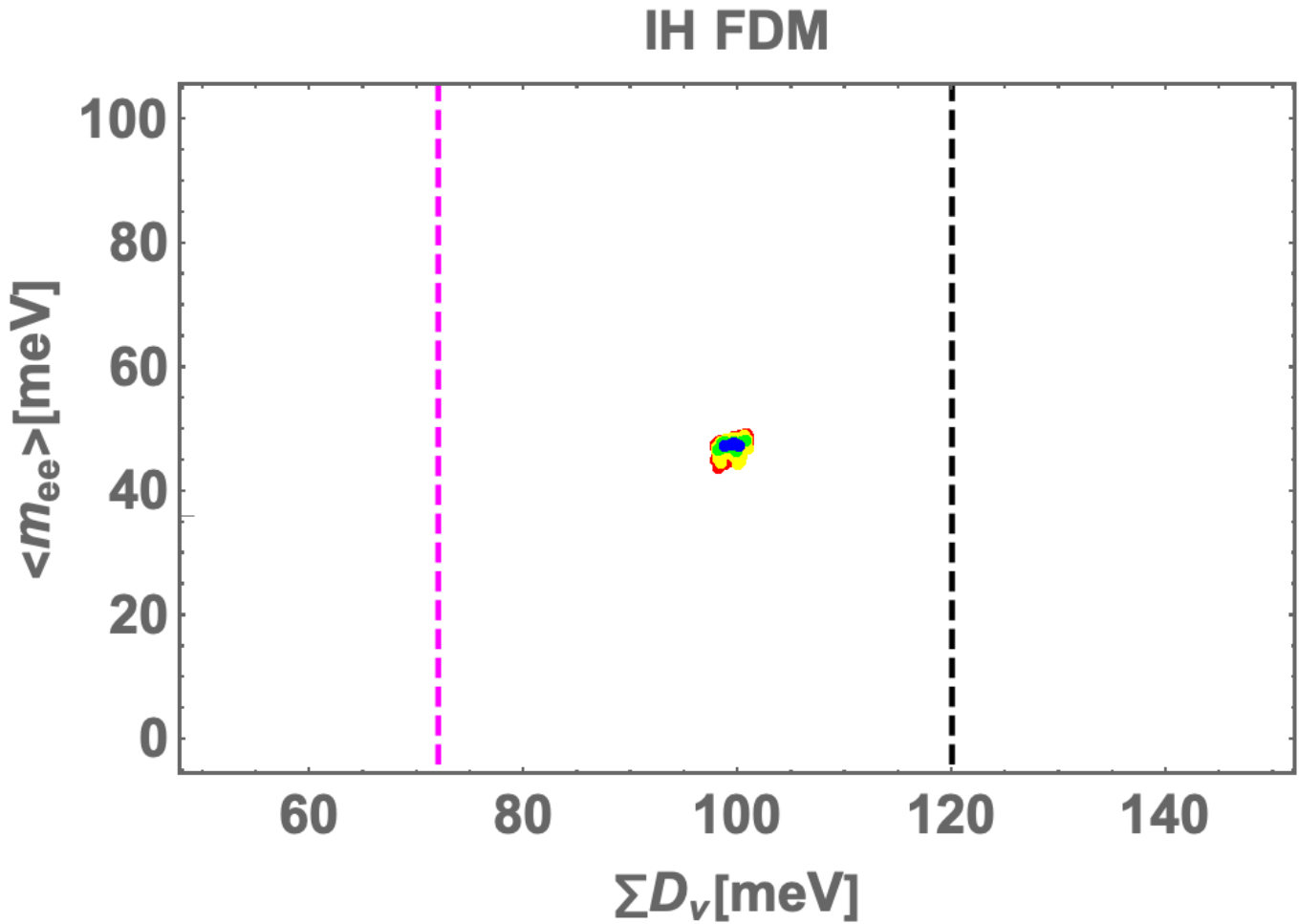}
\caption{Allowed regions for Majorana phase $\alpha_{21}$ in terms of $\delta_{\rm CP}$(up-left), $\langle m_{ee}\rangle$ in terms of $\delta_{\rm CP}$ (up-right), $\delta_{\rm CP}$ in terms of $\sum D_\nu$ (down-left), and  $\langle m_{ee}\rangle$ in terms of $\sum D_\nu$ (down-right). The color legends are the same as Fig.~\ref{fig:nhfdm1}. In down figures, the vertical dotted magenta lines are the experimental result  DESI and CMB data combination.}   
\label{fig:ihfdm2}\end{center}\end{figure}

Fig.~\ref{fig:ihfdm2} shows allowed regions for Majorana phase $\alpha_{21}$ in terms of $\delta_{\rm CP}$(up-left), $\langle m_{ee}\rangle$ in terms of $\delta_{\rm CP}$ (up-right), $\delta_{\rm CP}$ in terms of $\sum D_\nu$ (down-left), and  $\langle m_{ee}\rangle$ in terms of $\sum 
D_\nu$ (down-right). The color legends are the same as Fig.~\ref{fig:nhfdm1}. 
In down figures, the vertical dotted magenta lines are the experimental result  DESI and CMB data combination $\sum D_\nu\le$ 72 meV.
These figures predict as follows:
Majorana phases run over whole the regions but $\delta_{\rm CP}$ tends to be localized at nearby $50^\circ$ and $300^\circ$.
$\langle m_{ee}\rangle$ and $\sum D_\nu$ are respectively localized at nearby $50\ {\rm meV}$ and $100\ {\rm meV}$.

\begin{figure}[tb]\begin{center}
\includegraphics[width=80mm]{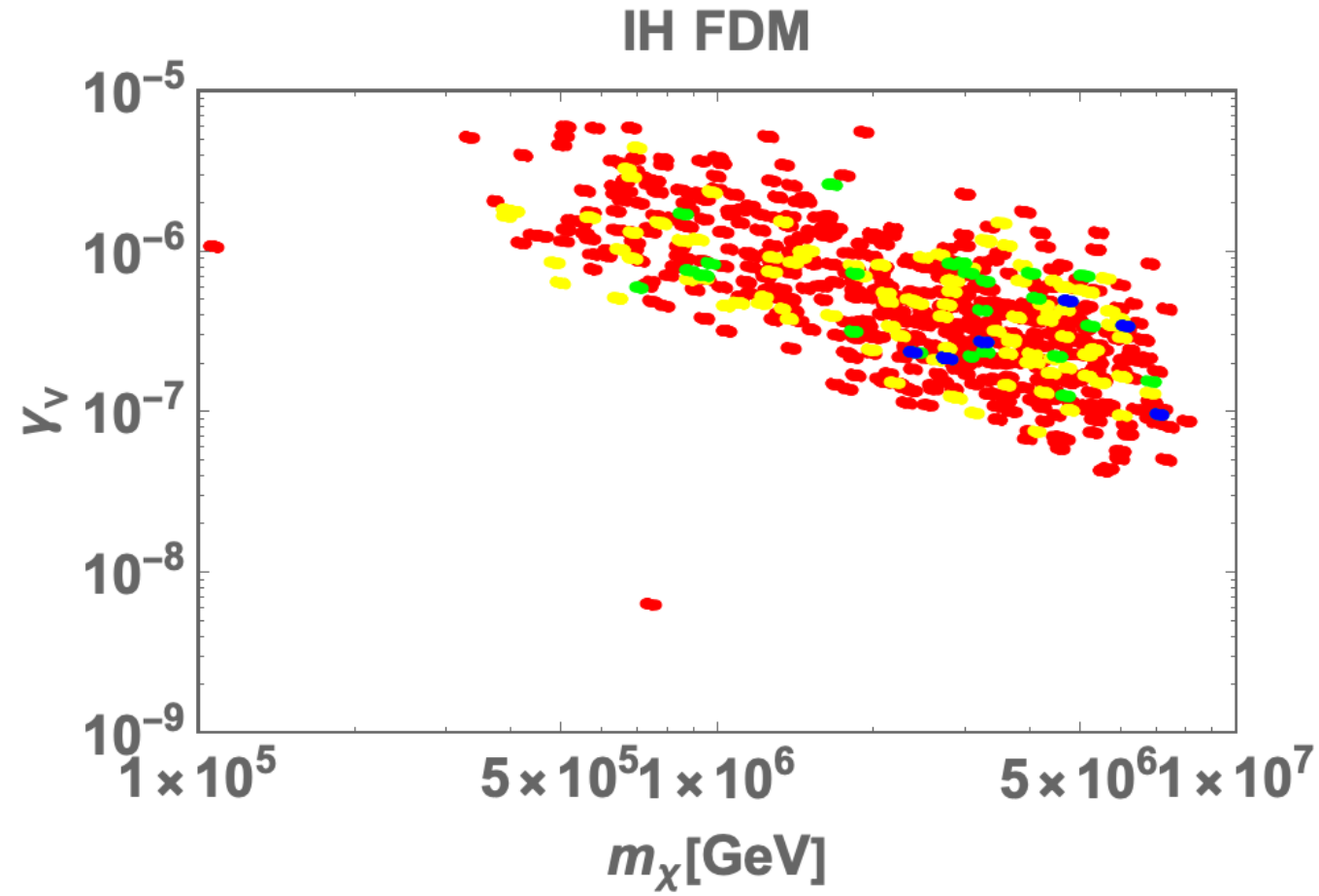}
\caption{Allowed region of $\gamma_\nu$ in terms of DM mass where the color legends are the same as Fig.~\ref{fig:nhfdm1}.}   
\label{fig:ihfdm3}\end{center}\end{figure}
Fig.~\ref{fig:ihfdm3} demonstrates allowed region of $\gamma_{\nu}$ in terms of $m_{\chi}$ where the color legends are the same as Fig.~\ref{fig:nhfdm1}.
The figure suggests that our DM mass range is ($10^5$, $10^7$) GeV and $\gamma_\nu$ is ($10^{-8}$, $10^{^5}$) with rather strong correlation.

In LFVs, we have obtained their maximum branching ratios $2.58\times10^{-39}$ for $\mu\to e\gamma$,
 $3.98\times10^{-39}$ for $\tau\to e\gamma$, and  $7.54\times10^{-38}$ for $\tau\to \mu\gamma$.
Thus, these branching ratios are far from the experimental bounds in Eq.~(\ref{eq:lfvs-cond}).


\subsubsection{Bosonic DM}

\begin{figure}[tb]\begin{center}
\includegraphics[width=53mm]{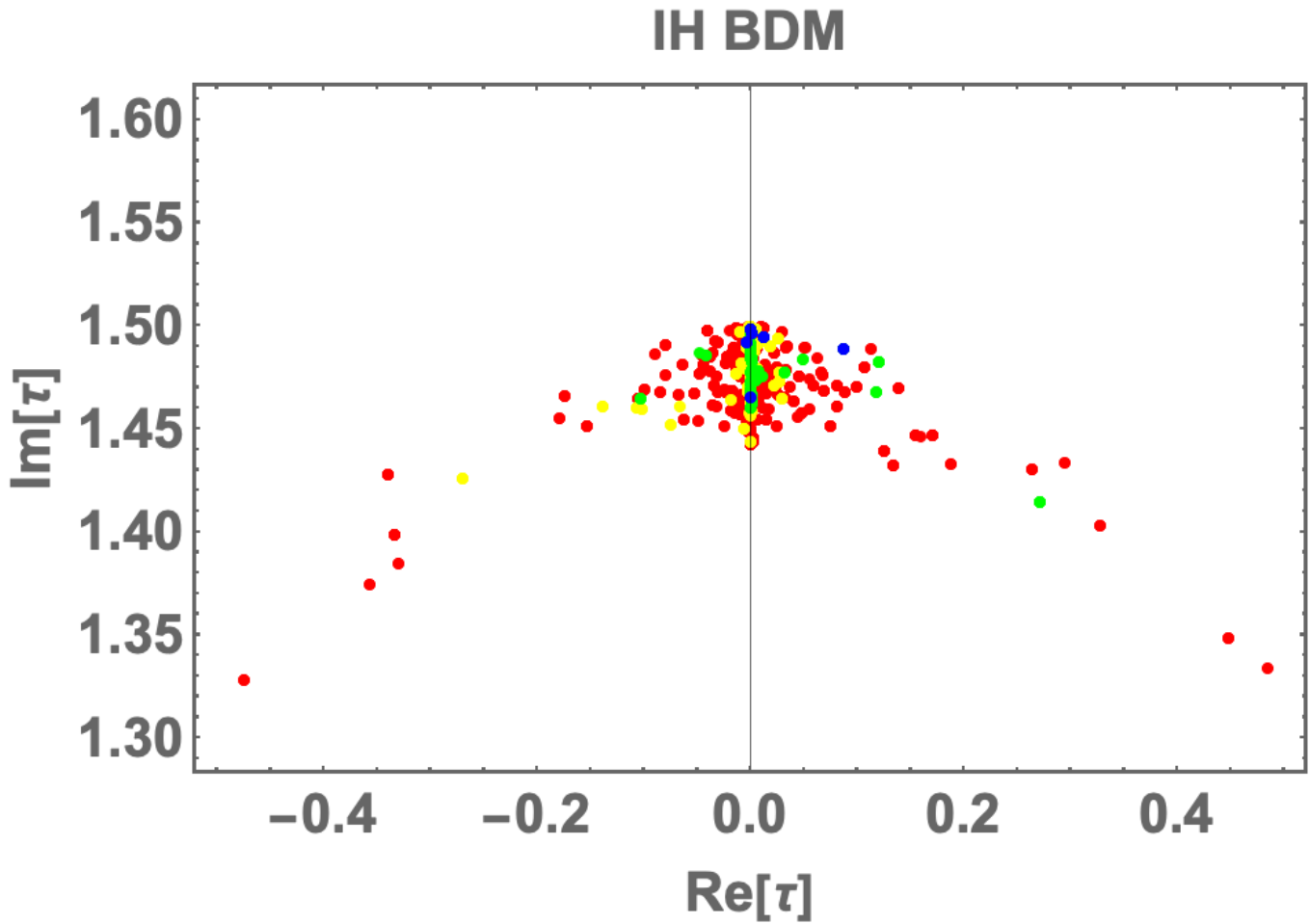}
\includegraphics[width=53mm]{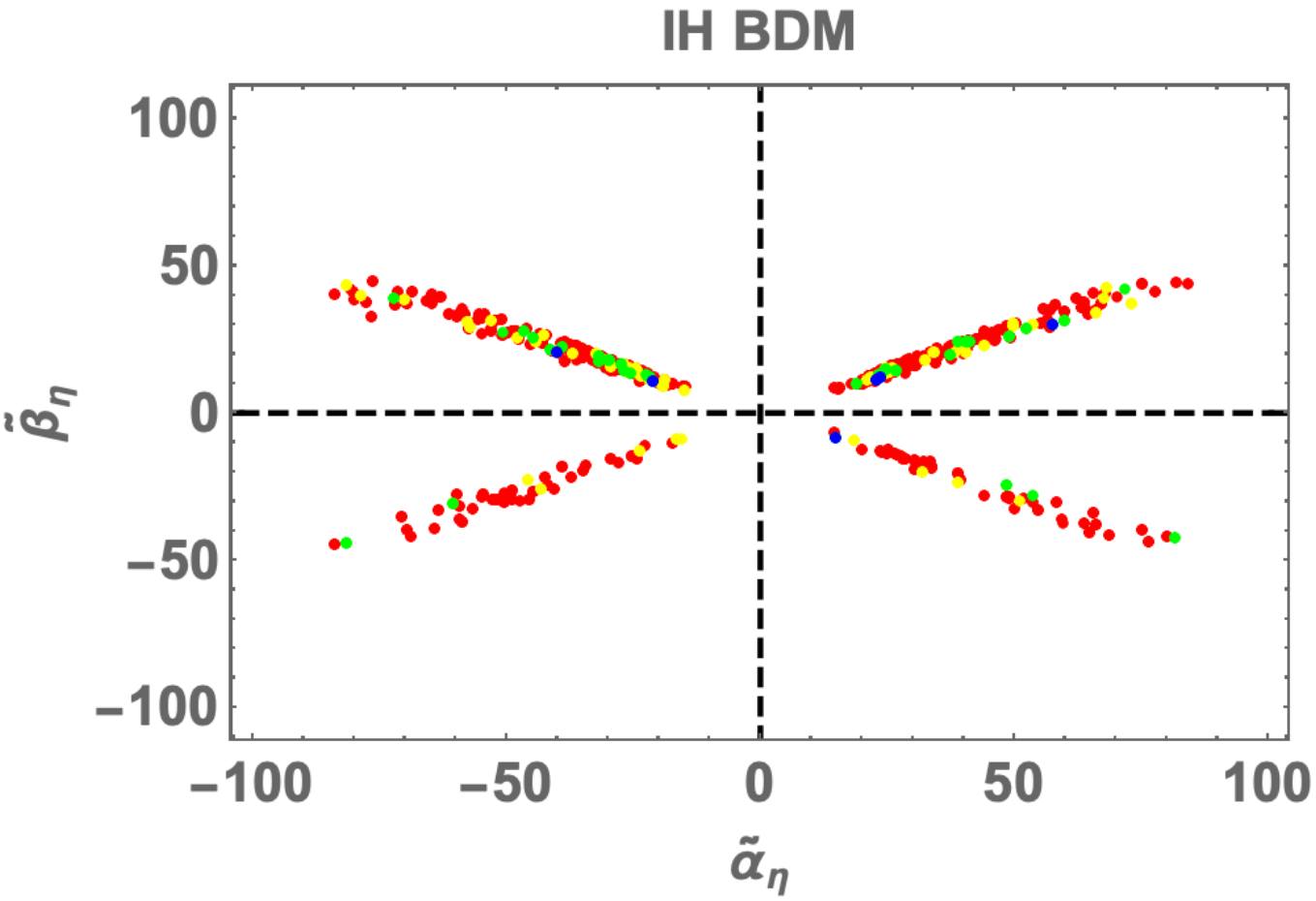}
\includegraphics[width=53mm]{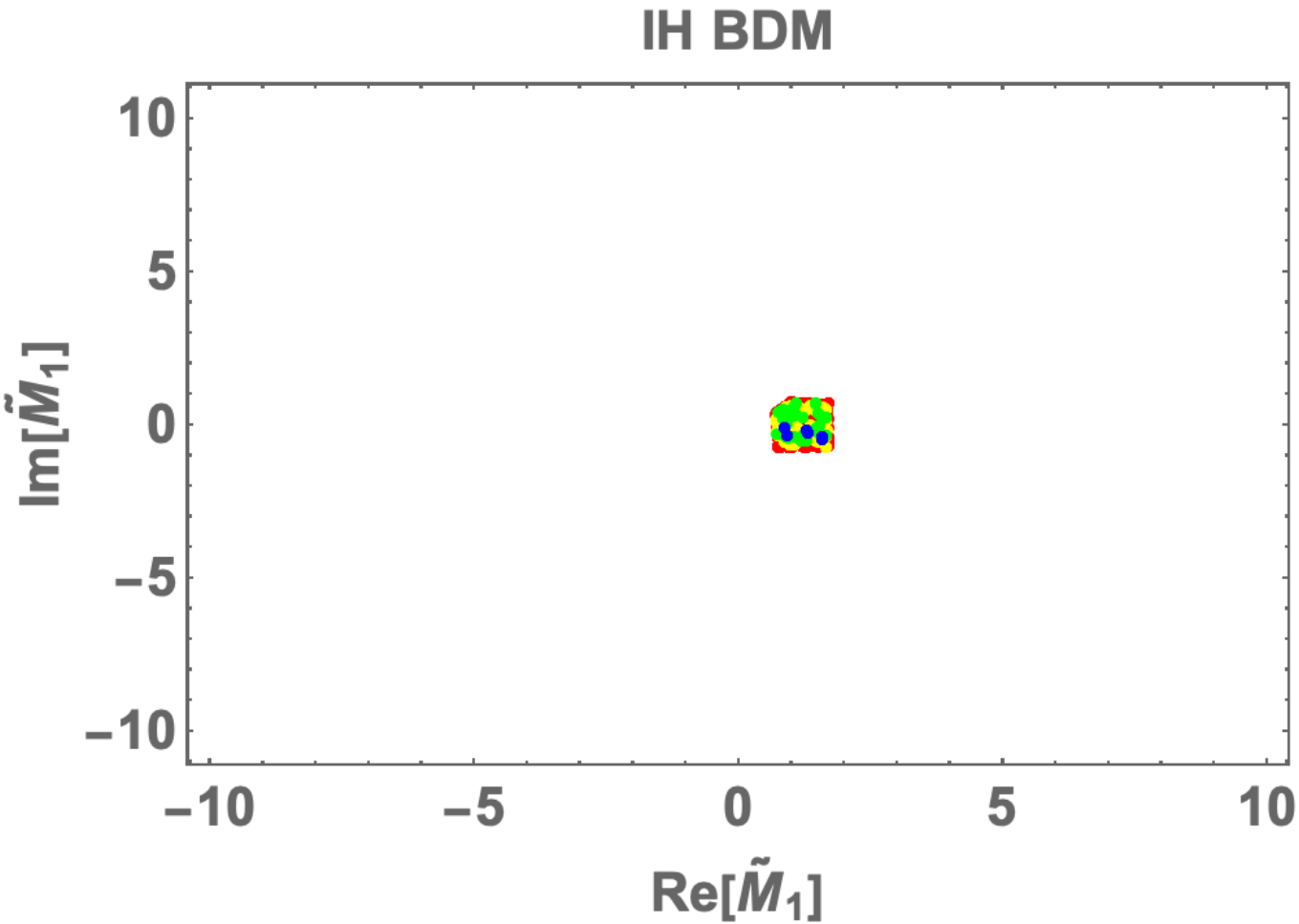}
\caption{Allowed regions for input parameters $\tau$(left-side), $\tilde\alpha$ and $\tilde\beta$(center), and $\tilde M_1$(right-side), where the color legends are the same as Fig.~\ref{fig:nhfdm1}. } 
\label{fig:ihbdm1}
\end{center}\end{figure}

In Fig.~\ref{fig:ihbdm1}, we show the allowed ranges for our input parameters; $\tau$(left-side), $\tilde\alpha$ and $\tilde\beta$(center), and $\tilde M_1$(right-side) where the color legends are the same as Fig.~\ref{fig:nhfdm1}.
The left-side figure tells us $|{\rm Re}[\tau]|\lesssim 0.5$ and $1.33\lesssim {\rm Im}[\tau]\lesssim 1.50$.
The center-side figure suggests $10\lesssim |\tilde\alpha|\lesssim 90$ and $5\lesssim \tilde\beta \lesssim 50$.
The right-side figure implies $0\lesssim {\rm Re}[\tilde M_1] \lesssim 2$ and $0\lesssim {\rm Im}[\tilde M_1]| \lesssim 1$.

\begin{figure}[tb]\begin{center}
\includegraphics[width=80mm]{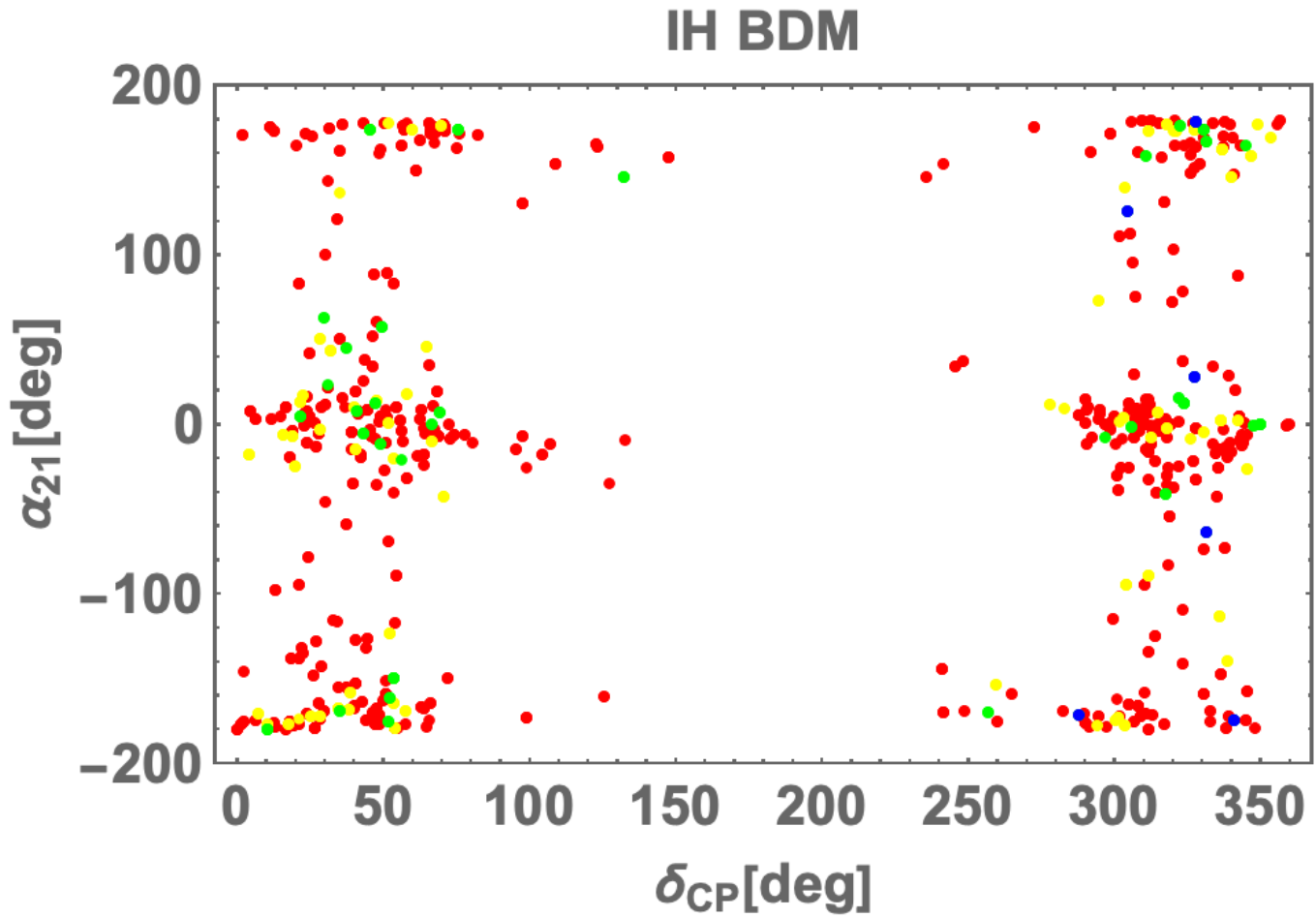}
\includegraphics[width=80mm]{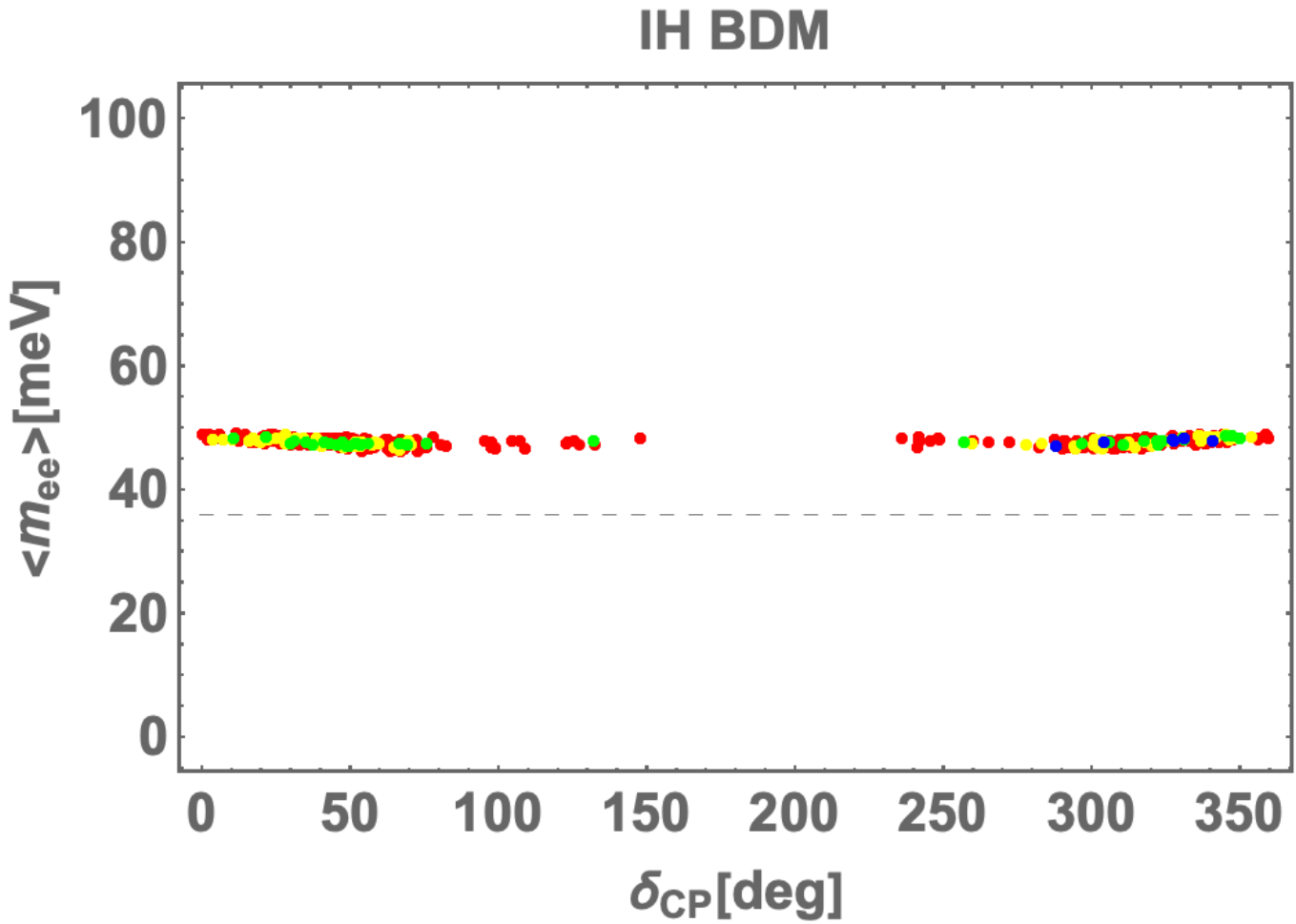}\\
\includegraphics[width=80mm]{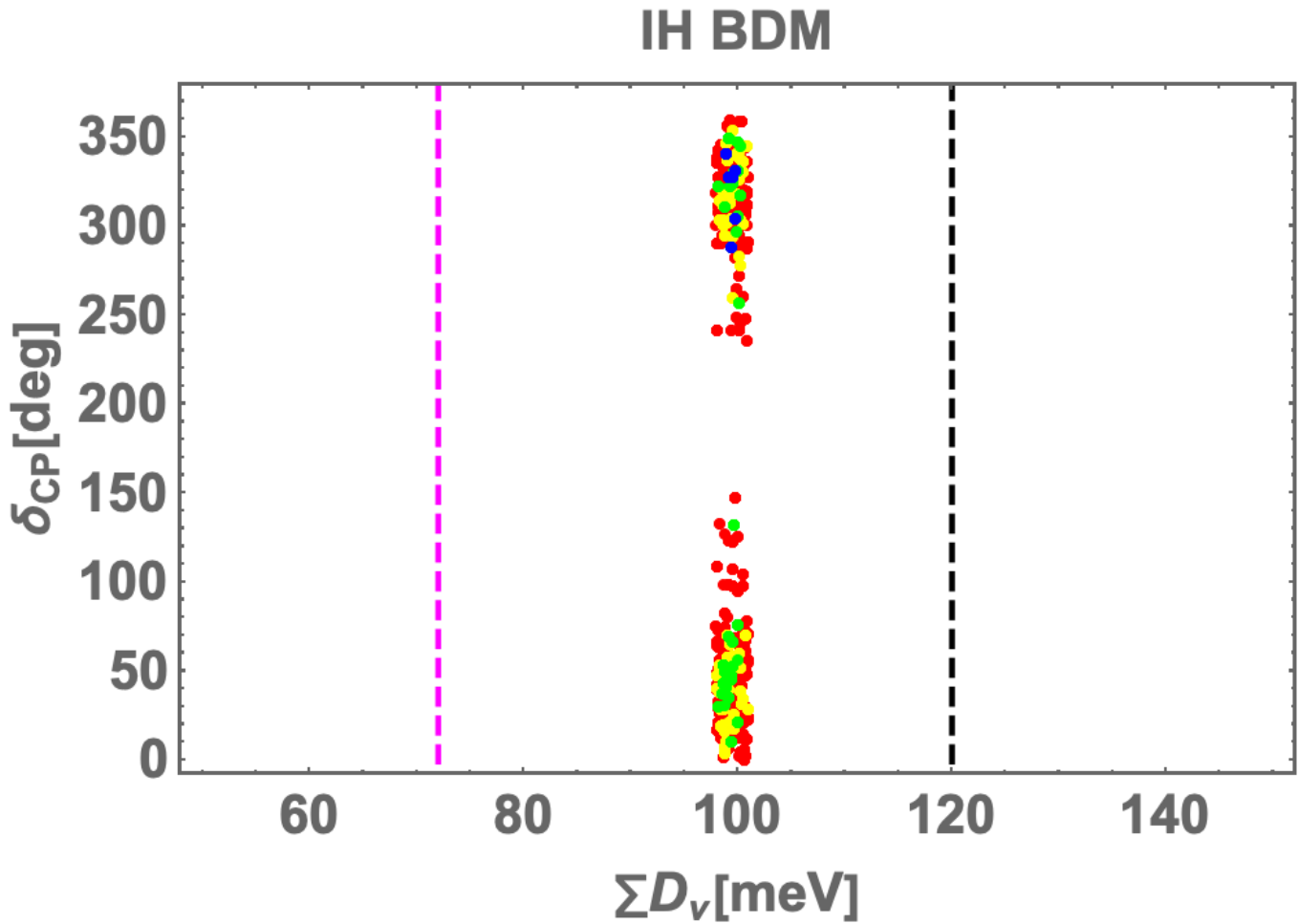}
\includegraphics[width=80mm]{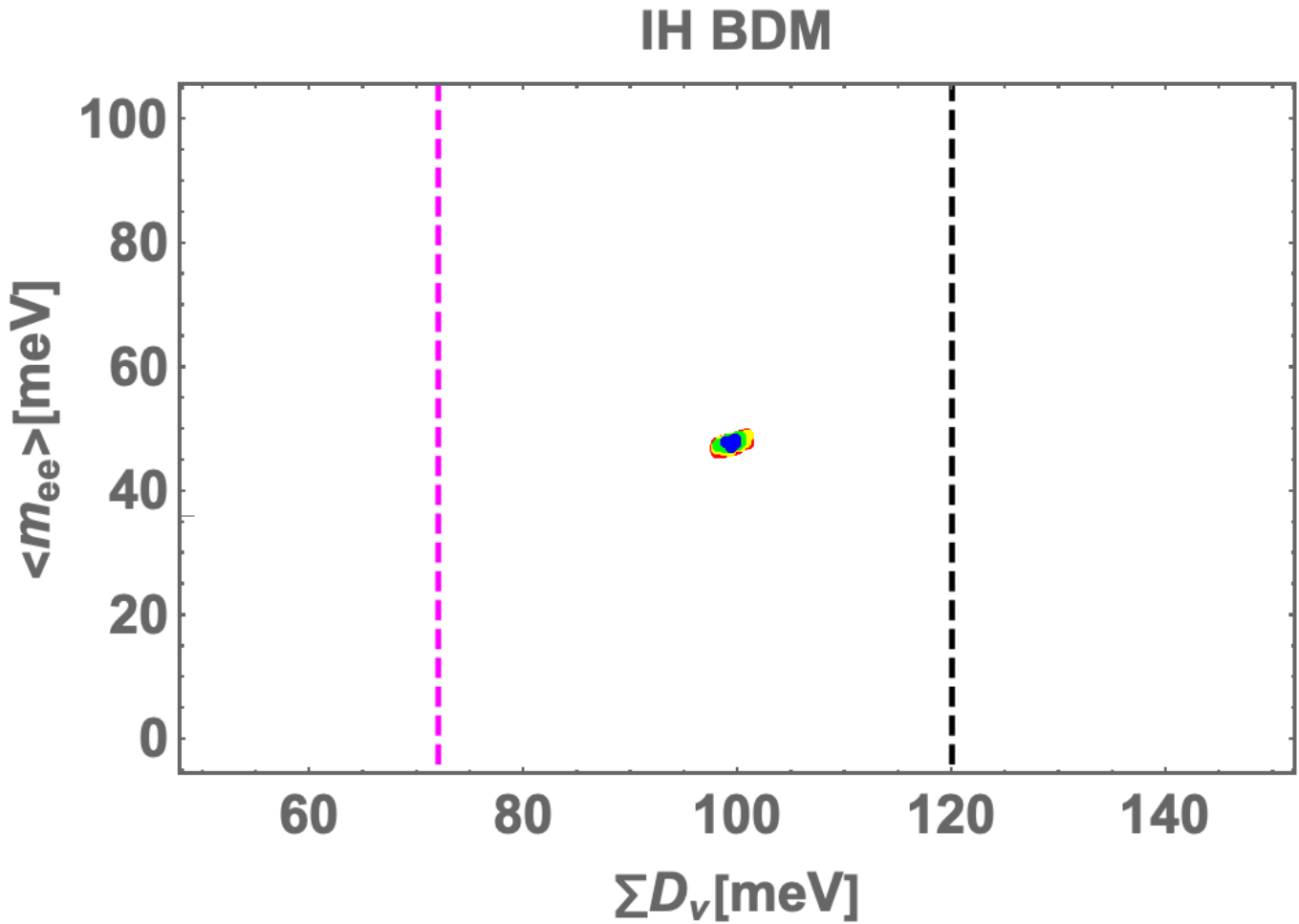}
\caption{Allowed regions for Majorana phase $\alpha_{21}$ in terms of $\delta_{\rm CP}$(up-left), $\langle m_{ee}\rangle$ in terms of $\delta_{\rm CP}$ (up-right), $\delta_{\rm CP}$ in terms of $\sum D_\nu$ (down-left), and  $\langle m_{ee}\rangle$ in terms of $\sum D_\nu$ (down-right). The color legends are the same as Fig.~\ref{fig:nhfdm1}. In down figures, the vertical dotted magenta lines are the experimental result  DESI and CMB data combination.}   
\label{fig:ihbdm2}\end{center}\end{figure}

Fig.~\ref{fig:ihbdm2} shows allowed regions for Majorana phase $\alpha_{21}$ in terms of $\delta_{\rm CP}$(up-left), $\langle m_{ee}\rangle$ in terms of $\delta_{\rm CP}$ (up-right), $\delta_{\rm CP}$ in terms of $\sum D_\nu$ (down-left), and  $\langle m_{ee}\rangle$ in terms of $\sum D_\nu$ (down-right). The color legends are the same as Fig.~\ref{fig:nhfdm1}. 
These figures predict to be $0^\circ\lesssim \delta_{\rm CP}\lesssim 150^\circ,\ 230^\circ\lesssim \delta_{\rm CP}\lesssim 360^\circ$ and  
$\alpha_{21}$ runs over the whole range but tends to be localized at nearby $0^\circ$ and $180^\circ$.
$\langle m_{ee}\rangle$ and $\sum D_\nu$ are respectively localized at nearby $50\ {\rm meV}$ and $100\ {\rm meV}$.

\begin{figure}[tb]\begin{center}
\includegraphics[width=80mm]{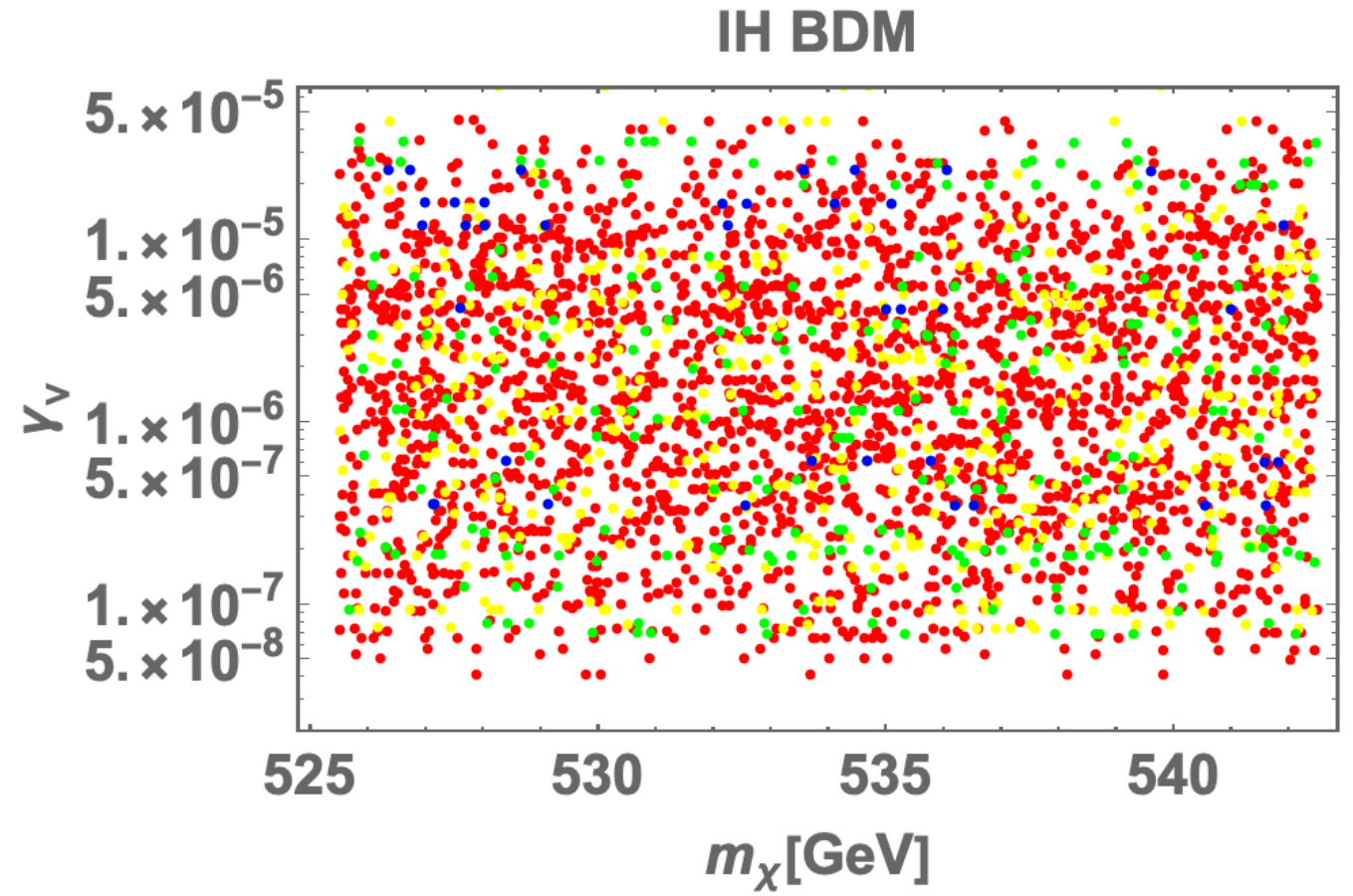}
\caption{Allowed region of $\gamma_\nu$ in terms of DM mass where the color legends are the same as Fig.~\ref{fig:nhfdm1}.}   
\label{fig:ihbdm3}\end{center}\end{figure}
Fig.~\ref{fig:ihbdm3} demonstrates allowed region of $\gamma_{\nu}$ in terms of $m_{\chi}$ where the color legends are the same as Fig.~\ref{fig:nhfdm1}.
The figure suggests $\gamma_\nu$ has to be small; $5\times10^{-8}-5\times 10^{-5}$, due to satisfying the constraints of LFVs.

In LFVs, we have obtained their maximum branching ratios $2.15\times10^{-22}$ for $\mu\to e\gamma$,
 $9.15\times10^{-25}$ for $\tau\to e\gamma$, and  $1.65\times10^{-24}$ for $\tau\to \mu\gamma$.
Thus, these branching ratios are far from the experimental bounds in Eq.~(\ref{eq:lfvs-cond}).

\section{Conclusion and discussion}
\label{sec:conclusion}
We have studied a non-holomorphic modular $S_3$ flavor symmetry in which we analyze neutrino sector, dark matter, and lepton flavor violations.
The active neutrino mass is generated via one-loop level.
We have performed chi-square analysis in the lepton sector and demonstrated some predictions in cases of normal hierarchy with fermionic or bosonic dark matter and inverted hierarchy of fermionic or bosonic dark matter, respectively.



\bibliography{ms3nh1lp.bib}

\end{document}